\documentclass[11pt,nofootinbib,showpacs]{revtex4}
\usepackage{amsfonts,amssymb,amsmath,graphicx,multirow}
\def\dac{\displaystyle\frac}

\def\[{\left[}
\def\]{\right]}
\def\({\left(}
\def\){\right)}

\newcommand{\diag}{\mathop{\rm diag}\nolimits}

\begin{document}

\baselineskip7mm
\title{Cosmological dynamics of spatially flat Einstein-Gauss-Bonnet models in various dimensions. Vacuum case}

\author{Sergey A. Pavluchenko}
\affiliation{Programa de P\'os-Gradua\c{c}\~ao em F\'isica, Universidade Federal do Maranh\~ao (UFMA), 65085-580, S\~ao Lu\'is, Maranh\~ao, Brazil}

\begin{abstract}

In this paper we perform a systematic study of vacuum spatially flat anisotropic $((3+D)+1)$-dimensional Einstein-Gauss-Bonnet cosmological models. We consider models which topologically are the product of 
two flat
isotropic submanifolds with different scale factors. One of these submanifolds is three-dimensional and represents
our 3D space and the other is
$D$-dimensional and represents extra dimensions. We consider no {\it ansatz} of the scale factors, which makes our results quite general.
With both Einstein-Hilbert and Gauss-Bonnet contributions in play and
with the symmetry involved, the cases with $D=1$, $D=2$, $D=3$ and $D\geqslant 4$ have different dynamics due to different structure of the equations of motion. We analytically
analyze equations of motion in all cases and describe all possible regimes. It appears that the only regimes with nonsingular future asymptotes are the Kasner regime in General Relativity as well as 
exponential regimes. As of
the past asymptotes, for a smooth transition only Kasner regime in Gauss-Bonnet is an option. With that at hand, we are down only to two viable regimes -- ``pure'' Kasner regime (transition from high-energy 
(Gauss-Bonnet) to
low-energy (General Relativity) Kasner regime) and a transition from high-energy
Kasner regime to anisotropic exponential solution. It appears that these regimes take place for different signs
of the Gauss-Bonnet coupling $\alpha$:
``pure'' Kasner regime occur for $\alpha > 0$ at low $D$ and $\alpha < 0$ for high $D$; anisotropic exponential regime is reached only for $\alpha > 0$. So if we restrain ourselves with $\alpha > 0$ 
solutions (that would be the case, say, if we identify $\alpha$ with inverse string tension in heterotic string theory), the only late-time regimes are Kasner for $D=1,\,2$ and anisotropic exponential
for $D\geqslant 2$.
Also, low-energy Kasner regimes ($a(t)\propto t^p$) have 
expansion rates for (3+1)-dimensional subspace (``our Universe'') ranging from $p=0.5$ ($D=1$) to $p=1/\sqrt{3} \approx 0.577$ ($D\to\infty$), which contradicts with dust-dominated Friedmann prediction ($p=2/3$).

\end{abstract}



\maketitle

\section{Introduction}

It is already more then hundred years to Einstein's General Relativity, but the extra-dimensional models are even older. Indeed,
the first attempt to construct extra-dimensional model was performed by Nordstr\"om~\cite{Nord1914} in 1914. It was a vector theory which unified Nordstr\"om's second gravity theory~\cite{Nord_2grav}
with Maxwell's electromagnetism. Back then Einstein's General Relativity (GR) has not been fully formulated yet and so it was natural that this kind of theory arises. Later in 1915 Einstein introduced his
theory~\cite{einst}, but still it took almost four years to prove that Nordstr\"om's theory was wrong. During Solar eclipse in 1919 there were performed measurements of light bending near the Sun and the 
deflection
angle was in perfect agreement with GR while Nordstr\"om's theory, being scalar gravity, predicted zeroth deflection angle.

Yet, the Nordstr\"om's idea on extra dimensions remained and in 1919 Kaluza proposed~\cite{KK1} similar model based on GR: in his model 5D Einstein equations could be decomposed into 4D Einstein equations
plus Maxwell's electromagnetism. In order to perform such decomposition, extra dimension should be ``curled'' or compactified into a circle and ``cylindrical conditions'' should be imposed. Later in 1926,
Klein introduced~\cite{KK2, KK3} nice quantum mechanical interpretation of this extra dimension and so the theory called Kaluza-Klein was formally formulated. Back then their theory unified all
known at that time interactions. With time, more interactions were known and it became clear that to unify them all more extra dimensions are needed. Nowadays, one of the promising theories to unify
all interactions is M/string theory.

Presence in the Lagrangian of the corrections which are squared in curvature is one of the distinguishing features of the gravitational counterpart of string theories.
Indeed, Scherk and Schwarz~\cite{sch-sch} were
first to discover the potential presence of the $R^2$ and
$R_{\mu \nu} R^{\mu \nu}$ terms in the Lagrangian of the Virasoro-Shapiro
model~\cite{VSh1, VSh2}. Curvature squared term of the $R^{\mu \nu \lambda \rho}
R_{\mu \nu \lambda \rho}$ type appears~\cite{Candelas_etal} in the low energy limit
of the $E_8 \times E_8$ heterotic superstring~\cite{Gross_etal} to match the kinetic term
for the Yang-Mills field. Later it was demonstrated~\cite{zwiebach} that the only
combination of quadratic terms that leads to ghost-free nontrivial gravitation
interaction is the Gauss-Bonnet (GB) term:

$$
L_{GB} = L_2 = R_{\mu \nu \lambda \rho} R^{\mu \nu \lambda \rho} - 4 R_{\mu \nu} R^{\mu \nu} + R^2.
$$

\noindent This term, first found
by Lanczos~\cite{Lanczos1, Lanczos2} (therefore it is sometimes referred to
as the Lanczos term) is an Euler topological invariant in (3+1)-dimensional
space-time, but not in (4+1) and higher dimensions.
Zumino~\cite{zumino} extended Zwiebach's result on higher
than squared curvature terms, supporting the idea that the low energy limit of the unified
theory might have a Lagrangian density as a sum of contributions of different powers of curvature. In this regard Einstein-Gauss-Bonnet (EGB) gravity could be seen as a subcase of more general Lovelock
gravity~\cite{Lovelock}, but in current paper we restrain ourselves with only quadratic corrections and so to EGB case.

Theories with extra dimensions have one thing in common -- one needs to explain where these additional dimensions are ``hiding'' in, as we do not sense them, at least with current level of experiments. 
One of
the ways to ``hide'' extra dimensions, as well as to recover 4D physics is to build so-called ``spontaneous compactification'' solution. Exact static solutions where the metric is a cross product of a
(3+1)-dimensional manifold and a constant curvature ``inner space'',  were discussed for the first time in~\cite{add_1}, but with (3+1)-dimensional manifold being actually Minkowski (the generalization for
a constant curvature Lorentzian manifold was done in~\cite{Deruelle2}).
In the context of cosmology, it is more interesting to consider a spontaneous compactification in the case where the four dimensional part is given by a Friedmann-Robertson-Walker metric.
In this case it is completely natural to consider also the size of the extra dimensions as time dependent rather then static. Indeed in
\cite{add_4} it was explicitly shown  that in order to have a more realistic model one needs to consider the dynamical evolution of the extra dimensional scale factor as well.
In~\cite{Deruelle2}, the equations of motion for compactification with both time dependent scale factors were written for arbitrary Lovelock order in the special case of spatially flat metric (the results 
were further proven in~\cite{prd09}).
The results of~\cite{Deruelle2} were reanalyzed for the special case of 10 space-time dimensions in~\cite{add_10}.
In~\cite{add_8}, the existence of dynamical compactification solutions was studied with the use of Hamiltonian formalism.
More recently, efforts on finding spontaneous  compactifications have been done in  \cite{add13} where
the dynamical compactification of (5+1) Einstein-Gauss-Bonnet model was considered, in \cite{MO04, MO14} with different metric {\it ansatz} for scale factors
corresponding to (3+1)- and extra dimensional parts, and in \cite{CGP1, CGP2, CGPT} where general (e.g. without any {\it ansatz}) scale factors and curved manifolds were considered. Also, apart from
cosmology, the recent analysis focuses on
properties of black holes in Gauss-Bonnet~\cite{addn_1, addn_2} and Lovelock~\cite{addn_3, addn_4} gravities, features of gravitational collapse in these
theories~\cite{addn_5, addn_6, addn_7}, general features of spherical-symmetric solutions~\cite{addn_8} and many others.

In the context of finding exact solutions, the most common {\it ansatz} used for the functional form of the scale factor is exponential or power law.
Exact solutions with exponential functions  for both  the (3+1)- and extra dimensional scale factors were studied for the first time in  \cite{Is86}, and exponentially increasing (3+1)-dimensional
scale factor and exponentially shrinking extra dimensional scale factor were described.
Power-law solutions have been analyzed  in \cite{Deruelle1, Deruelle2} and more  recently in~\cite{mpla09, prd09, Ivashchuk, prd10, grg10} so that there is an  almost complete description
(see also~\cite{PT} for useful comments regarding physical branches of the solutions).
Solutions with exponential scale factors~\cite{KPT} have been studied in detail, namely, models with both variable~\cite{CPT1} and constant~\cite{CST2} volume, developing a general scheme for
constructing solutions in EGB; recently~\cite{CPT3} this scheme
was generalized for general Lovelock gravity of any order and in any dimensions. Also, the stability of the solutions was addressed in~\cite{my15}, where it was
demonstrated that only a handful of the solutions could be called ``stable'' while the remaining are either unstable or have neutral/marginal stability and so additional investigation is
required.

In order to find all possible regimes of Einstein-Gauss-Bonnet cosmology, it is necessary to go beyond an exponential or power law {\it ansatz} and keep the functional form of the scale factor generic.
Of course in this case the equations of motion are much more complicated, but on the other hand, we are particularly interested in models which allow dynamical compactification, so it is natural to
consider metric as a product of spatially three-dimensional part and extra-dimensional. In that case three-dimensional part represents ``our Universe'' and we expect for this part to expand
while extra dimensional part should be suppressed in size with respect to three-dimensional one. In \cite{CGP1} it was found that there exists a phenomenologically
sensible regime in the case when the curvature of the extra dimensions is negative and the Einstein-Gauss-Bonnet theory does not admit a maximally symmetric solution. In this case the
three dimensional Hubble parameter and the extra dimensional scale factor asymptotically tend to the constant values. In \cite{CGP2} a detailed analysis of the cosmological dynamics in this model
with generic couplings was performed. Recently this model was also studied in~\cite{CGPT} where it was demonstrated that with an additional constraint on couplings Friedmann-type late-time behavior
could be restored.

In current paper, unlike~\cite{CGP1, CGP2, CGPT}, we consider both manifolds (three-dimensional and extra-dimensional) to be spatially flat and, similar to~\cite{CGP1, CGP2, CGPT}, put no {\it ansatz}
on the behavior the of scale factors; also, to be as general as possible, perform all the analysis analytically. In this paper we consider only vacuum model, so neither matter nor even a boundary term
(being just cosmological
or $\Lambda$-term in the absence of curvature) are considered -- we leave it to future consideration in a separate papers.
Of particular relevance to our present analysis
is~\cite{add13} where authors performed numerical analysis of 5D EGB model with $(3+2)$ splitting of the metric. Their approach was different from ours and so they have lost one of the branches while we
provide full analysis of the system.

The structure of the manuscript is as follows: first we write down general equations of motion for Einstein-Gauss-Bonnet gravity, then we rewrite them for our symmetry {\it ansatz}. In the following
sections we analyze them for $D=1$, $D=2$, $D=3$, and general $D\geqslant 4$ case, considering vacuum case in this paper only. Each case is
followed by a small discussion of the results and properties of this particular case; after considering all cases we discuss their properties, generalities and differences,
and draw conclusions.

\section{Equations of motion}

As mentioned above, we consider the spatially flat anisotropic cosmological model in Einstein-Gauss-Bonnet gravity without any matter source.
The equations of motion for such model include both first and second Lovelock contributions and could be easily derived from the general case (see e.g.~\cite{prd09}):

\begin{equation}
\begin{array}{l}
2 \[ \sum\limits_{j\ne i} (\dot H_j + H_j^2)
+ \sum\limits_{\substack{\{ k > l\} \\ \ne i}} H_k H_l \] + 8\alpha \[ \sum\limits_{j\ne i} (\dot H_j + H_j^2) \sum\limits_{\substack{\{k>l\} \\ \ne \{i, j\}}} H_k H_l +
3 \sum\limits_{\substack{\{ k > l >  \\   m > n\} \ne i}} H_k H_l
H_m H_n \] = 0
\end{array} \label{dyn_gen}
\end{equation}

\noindent as $i$th dynamical equation. The first Lovelock term -- Einstein-Hilbert contribution -- is in first squared parenthesis and the second term -- Gauss-Bonnet -- is in second parenthesis; $\alpha$
is the coupling constant for Gauss-Bonnet contribution and we put the corresponding constant for Einstein-Hilbert contribution to unity. Also, since we consider spatially flat cosmological model, scale
factors do not hold much physical sense and the equations are rewritten in terms of Hubble parameters $H_i = \dot a_i(t)/a_i(t)$. Apart from the dynamical equations we write down a constraint equation

\begin{equation}
\begin{array}{l}
2 \sum\limits_{i > j} H_i H_j + 24\alpha \sum\limits_{i > j > k > l} H_i H_j H_k H_l = 0.
\end{array} \label{con_gen}
\end{equation}

As mentioned in the Introduction,
we want to investigate the particular case with the scale factors splitted in two parts -- separately 3 dimensions (3-dimensional isotropic subspace), which are suppose to represent our world and the 
remaining represent the extra dimensions ($D$-dimensional isotropic subspace). So we put $H_1 = H_2 = H_3 = H$ and $H_4 = \ldots = H_{D+3} = h$ ($D$ designs the number of additional dimensions) and the
equations take the following form --
dynamical equation that corresponds to $H$:

\begin{equation}
\begin{array}{l}
2 \[ 2 \dot H + 3H^2 + D\dot h + \dac{D(D+1)}{2} h^2 + 2DHh\] + 8\alpha \[ 2\dot H \(DHh + \dac{D(D-1)}{2}h^2 \) + \right. \\ \\ \left. + D\dot h \(H^2 + 2(D-1)Hh + \dac{(D-1)(D-2)}{2}h^2 \) +
2DH^3h + \dac{D(5D-3)}{2} H^2h^2 + \right. \\ \\ \left. + D^2(D-1) Hh^3 + \dac{(D+1)D(D-1)(D-2)}{8} h^4 \] =0;
\end{array} \label{H_gen}
\end{equation}

\noindent the dynamical equation that corresponds to $h$:

\begin{equation}
\begin{array}{l}
2 \[ 3 \dot H + 6H^2 + (D-1)\dot h + \dac{D(D-1)}{2} h^2 + 3(D-1)Hh\] + 8\alpha \[ 3\dot H \(H^2 + 2(D-1)Hh + \right .\right. \\ \\ \left. \left. + \dac{(D-1)(D-2)}{2}h^2 \) +  (D-1)\dot h \(3H^2 + 3(D-2)Hh +
\dac{(D-2)(D-3)}{2}h^2 \) + 3H^4 +
\right. \\ \\ \left. + 9(D-1)H^3h + 3(D-1)(2D-3) H^2h^2 +  \dac{3(D-1)^2 (D-2)}{2} Hh^3 + \right. \\ \\ \left. + \dac{D(D-1)(D-2)(D-3)}{8} h^4 \] =0;
\end{array} \label{h_gen}
\end{equation}

\noindent and the constraint equation:

\begin{equation}
\begin{array}{l}
2 \[ 3H^2 + 3DHh + \dac{D(D-1)}{2} h^2 \] + 24\alpha \[ DH^3h + \dac{3D(D-1)}{2}H^2h^2 + \dac{D(D-1)(D-2)}{2}Hh^3 + \right. \\ \\ \left. + \dac{D(D-1)(D-2)(D-3)}{24}h^4\] = 0.
\end{array} \label{con2_gen}
\end{equation}

Looking at (\ref{H_gen}) and (\ref{h_gen}) one can see that for $D\geqslant 4$ the equations of motion contain the same terms, while for $D=\{1, 2, 3\}$ the terms are different (say, for $D=3$ terms with
$(D-3)$ multiplier are absent and so on) and so should be the dynamics.
We are going to study these four cases separately. As we mentioned in the Introduction, in this paper we are going to consider only vacuum case; the $\Lambda$-term case and possibly
general case with perfect fluid with arbitrary equation of state we as well as effect of curvature are going to be considered in the following papers.

\section{$D=1$ case}

In this case the equations of motion take form ($H$-equation, $h$-equation and constraint correspondingly):

\begin{equation}
\begin{array}{l}
4\dot H + 6H^2 + 2\dot h + 2h^2 + 4Hh + 8\alpha \( 2(\dot H + H^2)Hh + (\dot h + h^2)H^2\) = 0,
\end{array} \label{D1_H}
\end{equation}

\begin{equation}
\begin{array}{l}
6\dot H + 12H^2 + 24\alpha (\dot H + H^2)H^2 = 0,
\end{array} \label{D1_h}
\end{equation}

\begin{equation}
\begin{array}{l}
6H^2 + 6Hh + 24\alpha H^3h = 0.
\end{array} \label{D1_con}
\end{equation}

\begin{figure}
\includegraphics[width=1.0\textwidth, angle=0]{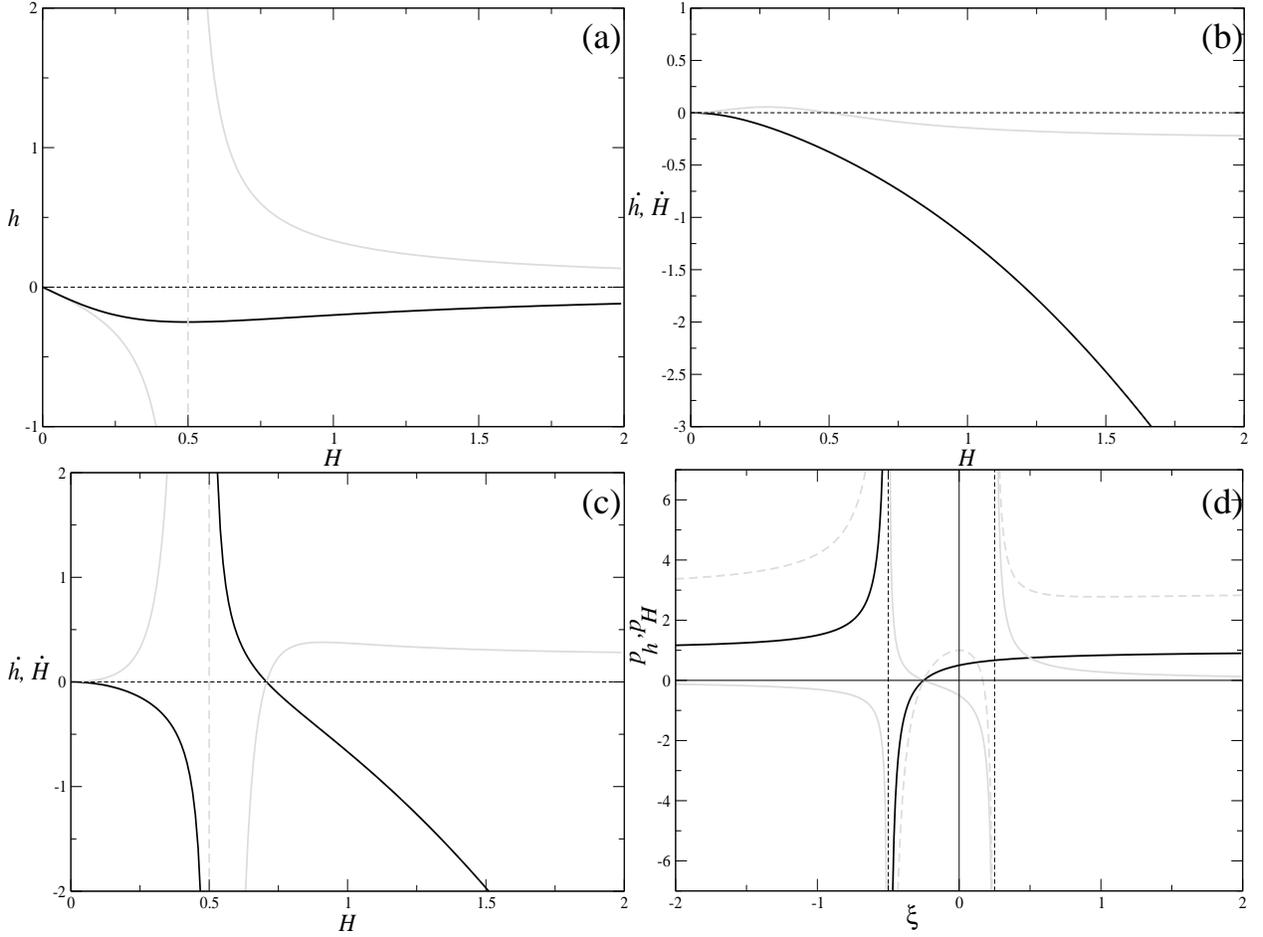}
\caption{Graphs illustrating the dynamics of $D=1$ vacuum cosmological model. In (a) panel we present the behavior for $h(H)$ from (\ref{D1_hh}) -- black for $\alpha > 0$ and grey for $\alpha < 0$. In panels
(b) and (c) we presented $\dot H(H)$ in black and $\dot h(H)$ in grey for $\alpha > 0$ ((b) panel) and $\alpha < 0$ ((c) panel). Finally in (d) panel we presented Kasner exponents $p_H$ in black, $p_h$ in
grey and the expansion rate $(3p_H + p_h)$ in dashed grey; irregularities denoted as dashed black lines (see the text for more details).}\label{D1L}
\end{figure}

From (\ref{D1_con}) we can easily see that

\begin{equation}
\begin{array}{l}
h = - \dac{H}{1+4\alpha H^2},
\end{array} \label{D1_hh}
\end{equation}

\noindent so that $H$ and $h$ always have opposite sign for $\alpha > 0$, but they could have same sign in $\alpha < 0$ case. We presented them in Fig.~\ref{D1L}(a) -- black for $\alpha > 0$ and grey for
$\alpha < 0$. Also one can resolve (\ref{D1_h}) for vacuum case with respect to
$\dot H$ to obtain

\begin{equation}
\begin{array}{l}
\dot H = - \dac{2H^2 (1+2\alpha H^2)}{1+4\alpha H^2};
\end{array} \label{D1_dH}
\end{equation}

\noindent after that with use of (\ref{D1_dH}) one can solve (\ref{D1_H}) to get

\begin{equation}
\begin{array}{l}
\dot h = - \dac{2H^2 (8\alpha^2 H^4 +  2\alpha H^2 - 1)}{(1+4\alpha H^2)(16\alpha^2 H^4 + 8\alpha H^2 + 1)}.
\end{array} \label{D1_dh}
\end{equation}

Now we can plot $\dot H$ and $\dot h$ versus $H$; we depicted them in Figs.~\ref{D1L}(b, c). Panel (b) corresponds to $\alpha > 0$ case and panel (c) -- to $\alpha < 0$; particular
curves correspond to $\alpha = \pm 1$.
In these panels in black we put $\dot H(H)$ and in grey -- $\dot h(H)$.

Now, let us handle non-singular asymptotic regimes for this case. From eqs. (\ref{D1_hh})--(\ref{D1_dh}) we can find that

\begin{equation}
\begin{array}{l}
\lim\limits_{H\to 0} \dac{h}{H} = -1, ~ \lim\limits_{H\to 0} \dac{\dot H}{H^2} = -2,~\lim\limits_{H\to \infty} \dac{h}{H} = 0, ~ \lim\limits_{H\to \infty} \dac{\dot H}{H^2} = -1.
\end{array} \label{D1_as1}
\end{equation}


The solution of the $\dot H/H^2 = k$ equation is $H(t) \propto -1/(kt)$, remembering the definition of the power-law {\it ansatz} $a(t) \propto t^{p}$ and comparing these two we find that $p = -1/k$,
so that for $H\to 0$ we have $p_H = 0.5$, $p_h = -0.5$ so that $\sum p_i = 3p_H + p_h = 1$ which corresponds to the Kasner solution in General Relativity. This result is quite expected --
from Fig. \ref{D1L}
we can see that this is late-time regime and, with Gauss-Bonnet contribution being second order on curvature, its effect on the dynamics wear off at late times. Hereafter we refer to this regime as $K_1$ --
Kasner regime with $\sum p_i = 1$.
Similarly, for $H\to\infty$ we have power-law behavior for both three-dimensional subspace with $p_H = 1$, and for extra-dimensional part with $p_h = 0$, so that we have Gauss-Bonnet Kasner regime with
$\sum p = 3$ and we denote it in a similar way as above -- $K_3$ -- Kasner regime with $\sum p = 3$.

There are no other regimes apart from these two for $\alpha > 0$ (see Fig.~\ref{D1L}(b)), but for $\alpha < 0$ there are. One cannot miss that $H\to 0$ asymptotes valid also for $\alpha < 0$ (see
Fig.~\ref{D1L}(c)), as well as for $H\to\infty$, so we denote them in the same way as in $\alpha > 0$ case.

The point $H^2_0 = - \dac{1}{4\alpha}$ for $\alpha < 0$ in vacuum case is physical singularity -- one can check that the equations of motion are discontinuous at that point and with $h$, $\dot h$ and $\dot H$
divergent the components of Riemann tensor are also divergent while $H$ remains regular. The last fact makes it similar to non-standard singularities, so we denote this regime as $nS$
(nonstandard singularity) and we will discuss them in the Discussion section.

Final asymptotic regime for vacuum case is the stable point $H \to H_1$ with $H_1^2 = - \dac{1}{2\alpha}$ (see Fig.~\ref{D1L}(c)). One can see from Eq. (\ref{D1_hh}) that $h(H_1) = H_1$ so that it is
isotropic solution. Also it gives $\dot H = \dot h \equiv 0$ with $H = h = \dac{1}{\sqrt{-2\alpha}}$ which correspond to the exponential solution; as expected, expressions for Hubble exponents coincide
with those obtained from exact solutions~\cite{CPT1}. We denote this regime as $E_{iso}$ -- as exponential isotropic solution. We summarize all regimes in Table \ref{D.1}.


\begin{table}[h]
\begin{center}
\caption{Summary of $D=1$ vacuum regimes.}
\label{D.1}
  \begin{tabular}{|c|c|c|}
    \hline
     $\alpha$  & Additional conditions & Regimes  \\
    \hline
 $\alpha > 0$  & no & $K_3 \to K_1$ \\  \cline{1-3}
  \multirow{3}{*}{$\alpha < 0$}  & $H^2 < - \frac{1}{4\alpha}$ & $nS\to K_1$  \\ \cline{2-3}
   & $- \frac{1}{2\alpha} > H^2 > - \frac{1}{4\alpha}$  & $nS\to E_{iso}$ \\  \cline{2-3}
   & $H^2 > - \frac{1}{2\alpha}$ & $K_3 \to E_{iso}$ \\
    \hline
  \end{tabular}
\end{center}
\end{table}

Finally it could be useful to rewrite equations (\ref{D1_dH}) and (\ref{D1_dh}) in terms of Kasner exponents -- from power-law {\it ansatz} $a(t) = a_0 t^p$ we can derive $p = - H^2/\dot H$ and so retrieve
expressions for $p_H$ and $p_h$ -- Kasner exponents associated with three- and extra-dimensional parts respectively:

\begin{equation}
\begin{array}{l}
p_H = \dac{1}{2}\times\dac{4\xi + 1}{2\xi + 1},~p_h = \dac{1}{2}\times\dac{4\xi + 1}{8\xi^2 + 2\xi - 1}~\mbox{with}~\xi = \alpha H^2.
\end{array} \label{D1_pHph}
\end{equation}

In this notation both Kasner exponents depends only on one variable $\xi$ whose sign indicate the sign of $\alpha$. We presented (\ref{D1_pHph}) in Fig.~\ref{D1L}(d) -- from it one can verify asymptotes
for both $H\to 0$ and $H\to\infty$. We depicted $p_H$ as black line, $p_h$ as solid grey line, $\sum p = 3p_H + p_h$ as dashed grey line; dashed black line corresponds to irregularities. From Fig.~\ref{D1L}(d)
one can clearly see that for $H\to\infty$ we have $p_H\to 1$ and $p_h\to 0$ and $\sum p \to 3$ as a result as well as $p_H = 0.5$ with $p_h = -0.5$ and so $\sum p = 1$ at $H=0$. As of irregularities,
$\xi = - 0.5$ correspond to the described above isotropic exponential solution (point $E_{iso}$ -- indeed, for exponential solutions Kasner exponents diverge (see, e.g.~\cite{PT} for the discussion of exponential
and power-law solutions and their relations),
while $\xi = 0.25$ is a regular point -- we can see that $p_H$ is regular there but $p_h$ is divergent -- it is caused by $\dot h = 0$ at that point (see Fig.~\ref{D1L}(b)). Finally, physical nonstandard
singularity depicted as $p_H = p_h = 0$ at $\xi = - 1/4$.

To conclude, in $D=1$ vacuum case there are total four regimes but only two of them are nonsingular -- $K_3 \to K_1$ for $\alpha > 0$ and $K_3 \to E_{iso}$ for $\alpha < 0$. Of these two only one
could be called viable --  $K_3 \to K_1$ for $\alpha > 0$ -- since the other one suppose isotropisation of the entire space and this is not what we observe.

\section{$D=2$ case}

In this case the equations of motion take form ($H$-equation, $h$-equation and constraint correspondingly):

\begin{equation}
\begin{array}{l}
4\dot H + 6H^2 + 4\dot h + 6h^2 + 8Hh + 8\alpha \( 2(\dot H + H^2) (2Hh + h^2) + 2(\dot h + h^2) (H^2 + 2Hh) +3H^2h^2\)  = 0,
\end{array} \label{D2_H}
\end{equation}

\begin{equation}
\begin{array}{l}
6\dot H + 12H^2 + 2\dot h + 2h^2 + 6Hh + 8\alpha \( 3(\dot H + H^2) (H^2 + 2Hh) + 3(\dot h + h^2)H^2 + 3H^3h  \)  = 0,
\end{array} \label{D2_h}
\end{equation}

\begin{equation}
\begin{array}{l}
6H^2 + 12Hh + 2h^2 + 24\alpha (2H^3h + 3H^2h^2 ) = 0.
\end{array} \label{D2_con}
\end{equation}

\begin{figure}
\includegraphics[width=1.0\textwidth, angle=0]{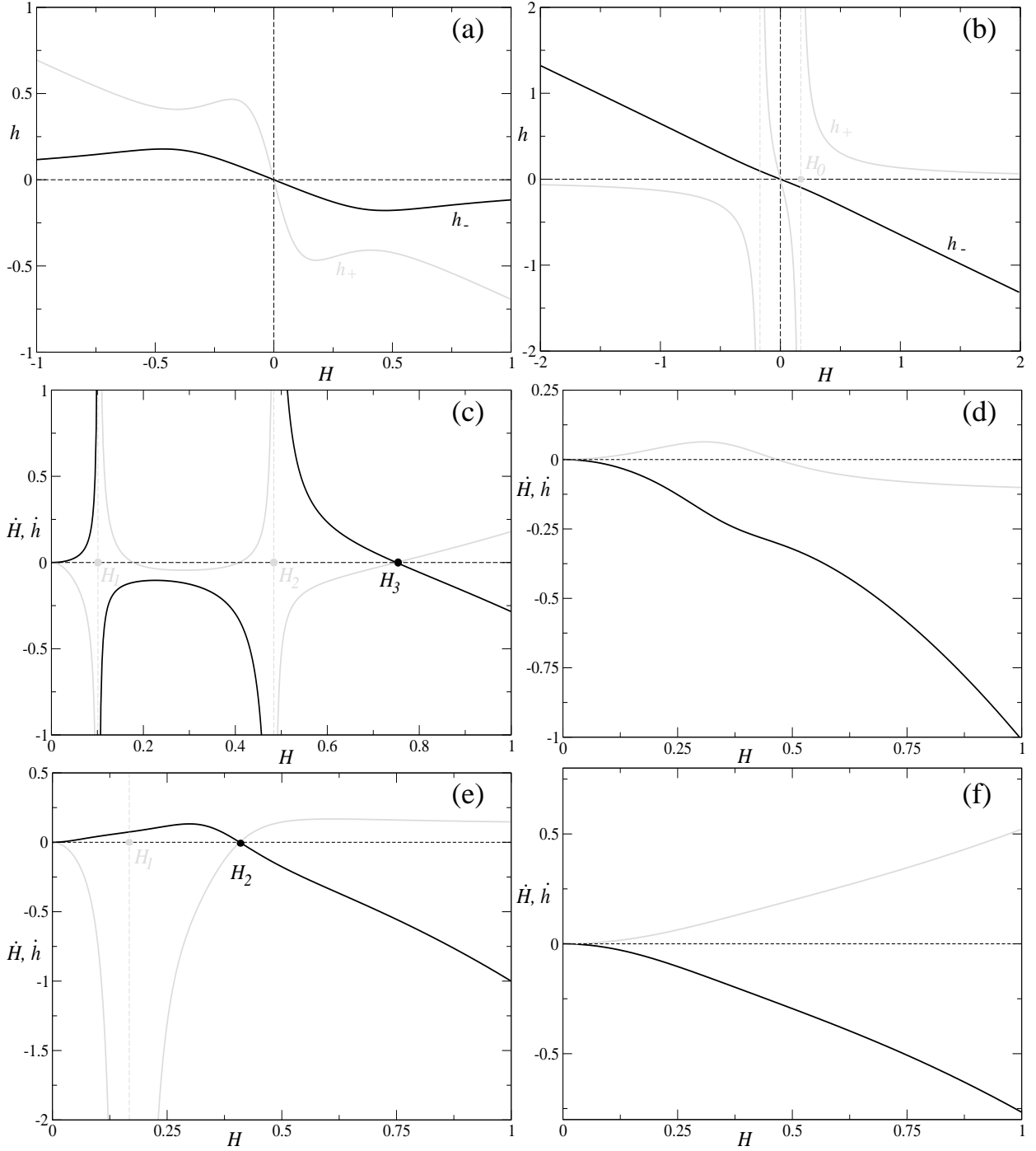}
\caption{Dynamics of $D=2$ vacuum model. In (a) and (b) panels we presented $h(H)$ curve for $\alpha > 0$ (a) and $\alpha < 0$ (b); both $h_\pm$ branches are presented (see (\ref{D2_hvac})).
In panels (c)--(f) we presented $\dot H(H)$ curves in black and $\dot h(H)$ curves in grey for the following cases: $\alpha > 0$, $h=h_+$ (c) panel, $\alpha > 0$, $h=h_-$ (d) panel,
$\alpha < 0$, $h=h_+$ (e) panel, $\alpha < 0$, $h=h_-$ (f) panel (see text for details).}\label{D2v}
\end{figure}

If we solve (\ref{D2_con}) with respect to $h$ we get

\begin{equation}
\begin{array}{l}
h_\pm = - \dac{H \( 3+12\alpha H^2 \pm \sqrt{6-36\alpha H^2 + 144\alpha^2 H^4} \)}{1+36\alpha H^2};
\end{array} \label{D2_hvac}
\end{equation}

\noindent one can see that the radicand is always positive and (one can easily verify it) both roots for $h$ have different sign from $H$ in $\alpha > 0$ case (see Fig. \ref{D2v}(a)).
One cannot also miss that the $\alpha < 0$ case, presented in Fig. \ref{D2v}(b), has singularity at $H_0 = \pm \sqrt{-\dac{1}{36\alpha}}$.

Now we can solve (\ref{D2_H})--(\ref{D2_h}) with respect to $\dot H$ and $\dot h$, substitute (\ref{D2_hvac}) to get $\dot H(H)$ and $\dot h(H)$ curves. The expressions for $\dot H(H)$ and
$\dot h(H)$ are as follows:

\begin{equation}
\begin{array}{l}
\dot H_\mp = - \dac{H^2 P_1^\pm}{Q^\pm}, \\
\dot h_\mp = \dac{3H^2}{(1+36\xi)^2} \dac{P_2^\pm}{Q^\pm}~\mbox{with}~\xi=\alpha H^2~\mbox{and}~\mathcal{D} = \sqrt{6-36\xi+144\xi^2}, \\
P_1^\mp = 24192\xi^4 - 4896\xi^3 \pm 575\mathcal{D}\xi^3 + 648\xi^2 \pm 624\mathcal{D}\xi^2 + 66\xi \pm 32\mathcal{D}\xi - 3 \pm 2\mathcal{D}, \\
P_2^\mp = 2488320\xi^6 - 2446848\xi^5 \mp 207360\mathcal{D}\xi^5 + 145152\xi^4 \pm 38016\mathcal{D}\xi^4 + 39744\xi^3 \pm \\ \pm 6048\mathcal{D}\xi^3 + 168\xi^2 \pm 648\mathcal{D}\xi^2 + 266\xi
\mp 80\mathcal{D}\xi - 7 \pm 3\mathcal{D}, \\
Q^\mp = 31104\xi^4 - 2880\xi^3 + 216\xi^2 \pm 384\mathcal{D}\xi^2 - 12\xi \pm 32\mathcal{D}\xi + 1.
\end{array} \label{D2_dHh}
\end{equation}

Now we can plot $\dot H(H)$ and $\dot h(H)$ curves, we presented them in Fig. \ref{D2v}(c)--(f). In this figure we presented $\dot H(H)$ curves in black and $\dot h(H)$ curves in grey
and the cases are the following: $\alpha > 0$, $h=h_+$ (c) panel, $\alpha > 0$, $h=h_-$ (d) panel, $\alpha < 0$, $h=h_+$ (e) panel, $\alpha < 0$, $h=h_-$ (f) panel. Before having closer look on the
panels, it is useful to find zeros of $\dot H$ and $\dot h$ from expressions in~(\ref{D2_dHh}):

\begin{equation}
\begin{array}{l}
P_1^+ = 0 \Leftrightarrow \xi = - \dac{1}{6},~\xi = \xi_0 = \dac{\sqrt[3]{10}}{9} + \dac{\sqrt[3]{100}}{36} + \dac{7}{36} \approx 0.56276; \\ \\
P_1^- = 0 \Leftrightarrow \xi = - \dac{1}{36}; \\ \\
P_2^+ = 0 \Leftrightarrow \xi = \pm \dac{1}{6},~\xi = \dac{1}{8} - \dac{\sqrt{5}}{24} \approx 0.03183,~\xi = \xi_0; \\ \\
P_2^- = 0 \Leftrightarrow \xi = - \dac{1}{36},~\xi = \dac{1}{8} + \dac{\sqrt{5}}{24} \approx 0.21817.
\end{array} \label{D2_roots}
\end{equation}

Also it is useful to find vertical asymptotes for $\dot H$ and $\dot h$:

\begin{equation}
\begin{array}{c}
Q^+ = 31104\xi^4 - 2880\xi^3 + 216\xi^2 - 384\mathcal{D}\xi^2 - 12\xi - 32\mathcal{D}\xi + 1 = 0 \\
\Updownarrow \\
11664 Z^6 - 5616 Z^5 + 612 Z^4 - 72 Z^3 + 12 Z^2 - 48 Z + 1 = 0~\mbox{with}~Z=\xi/2,\\
Z_1 \approx 0.0209 \Rightarrow \xi_1 \approx 0.0105,~Z_2 \approx 0.4690 \Rightarrow \xi_2 \approx 0.2345; \\
Q^- = 0 \Leftrightarrow \xi = - \dac{1}{36}.
\end{array} \label{D2_Qplus}
\end{equation}

Now let us have a closer look on the panels.
The curves in (c) panel ($\alpha > 0$, $h=h_+$) have two vertical asymptotes determined by $Q^+$ (\ref{D2_Qplus}), so in Fig. \ref{D2v}(c) $H_1$ corresponds to $\xi_1$ and $H_2$ -- to $\xi_2$.
From (c) panel one can clearly see that $H_1$ is singular attractor -- indeed, for $H_1 > H > 0$ we have positive
$\dot H$ so that $H\to H_1$ and for $H_2 > H > H_1$ we have negative $\dot H$ so that $H\to H_1$ again, and $H=H_1$ is singular. So that the only nonsingular regime exists for $H > H_2$ and this
regime is exponential and anisotropic: $h/H \approx -0.722$; its location is defined by positive root of $P_1^+$ (see~(\ref{D2_roots})); to get explicit value for $h/H$ ratio one can substitute
$H = \sqrt{\xi_0/\alpha}$ into $h_+$ (\ref{D2_hvac}).

The curves in (d) ($\alpha > 0$, $h=h_-$) and (f) ($\alpha < 0$, $h=h_-$) panels behave differently at early times but have the same asymptote at late times: $H,\,h \to 0$, and they reach
it as $h_-$ branch:

\begin{equation}
\begin{array}{c}
\lim\limits_{H\to 0} \dac{h_\pm}{H} = - 3 \mp \sqrt{6} < 0.
\end{array} \label{D2_vaclate}
\end{equation}

Finally in (e) panel of Fig. \ref{D2v} we presented the case $\alpha < 0$, $h=h_+$. One cannot miss singular behavior of $\dot h$ at $H=H_1$ while $\dot H$ is regular in this point. This is the same
singularity we saw in Fig. \ref{D2v}(b) at $H=H_0$. So we have $H=H_2$ as a stable point and it corresponds to exponential isotropic solution, unlike situation
with $\alpha > 0$. For $0 < H < H_1$ we face singularity at
$H=H_1$, so that isotropisation is reached only for $H > H_1$.  The value for $H_2$ is defined by roots of $P_1^+$ (see (\ref{D2_roots})).

Now let us address non-singular asymptotic regimes in this case. Similar to the previous section let us find the corresponding limits

\begin{equation}
\begin{array}{c}
\lim\limits_{H\to \infty} \dac{h}{H} = \dac{\pm\sqrt{\alpha^2}-\alpha}{3\alpha};\quad \lim\limits_{H\to 0} \dac{\dot H_\pm}{H^2} = 3 \mp 2\sqrt{6};\quad \lim\limits_{H\to \infty}
\dac{\dot H_\pm}{H^2} =  \dac{\pm 2\sqrt{\alpha^2} - 7\alpha}{9\alpha},
\end{array} \label{D2_vaclimits}
\end{equation}

\noindent and the last limit is (\ref{D2_vaclate}). From all four we can recover power-law behavior for both $h_\pm$ branches in both $H\to 0$ and $H\to\infty$ limits; we present them in Table
\ref{D.21}.

\begin{table}[h]
\begin{center}
\caption{Power-law behavior in $D=2$}
\label{D.21}
  \begin{tabular}{|c|c|c|c|c|c|}
    \hline
     $h\pm$ & $\alpha$   & $\lim H$ & $p_H$ & $p_h$ & $\sum p_i$   \\
    \hline
\multirow{4}{*}{$h_+$} & \multirow{2}{*}{$\alpha > 0$} & 0        & $- \frac{1}{3+2\sqrt{6}}$ & $\frac{2\sqrt{6}+5}{7+3\sqrt{6}}$ & 1 \\  \cline{3-6}
                       &                               & $\infty$ & $\frac{9}{5}$             & $- \frac{6}{5}$                  & 3 \\  \cline{2-6}
                       & \multirow{2}{*}{$\alpha < 0$} & 0        & $- \frac{1}{3+2\sqrt{6}}$ & $\frac{2\sqrt{6}+5}{7+3\sqrt{6}}$ & 1 \\  \cline{3-6}
                       &                               & $\infty$ & 1                        & 0                               & 3 \\  \cline{1-6}
\multirow{4}{*}{$h_-$} & \multirow{2}{*}{$\alpha > 0$} & 0        & $\frac{1}{2\sqrt{6}-3}$ & $\frac{2\sqrt{6}-5}{3\sqrt{6}-7}$ & 1 \\  \cline{3-6}
                       &                               & $\infty$ & 1                        & 0                               & 3 \\  \cline{2-6}
                       & \multirow{2}{*}{$\alpha < 0$} & 0        & $\frac{1}{2\sqrt{6}-3}$ & $\frac{2\sqrt{6}-5}{3\sqrt{6}-7}$ & 1 \\  \cline{3-6}
                       &                               & $\infty$ & $\frac{9}{5}$             & $- \frac{6}{5}$                  & 3 \\
\hline
  \end{tabular}
\end{center}
\end{table}

Finally, similar to the previous section it is useful to write down explicit expressions for Kasner exponents and plot them. As we noted in the previous section, Kasner exponent could be expressed in terms
of Hubble and its derivative as $p = - H^2/\dot H$, then with use of (\ref{D2_hvac}) and (\ref{D2_dHh}) we can obtain

\begin{equation}
\begin{array}{c}
p_H^\pm = \dac{Q^\pm}{P_1^\pm},~p_h^\pm = - \dac{(12\xi + 3 \pm \mathcal{D})^2 Q^\pm}{3P_2^\pm}.
\end{array} \label{D2_p}
\end{equation}

\begin{figure}
\includegraphics[width=1.0\textwidth, angle=0]{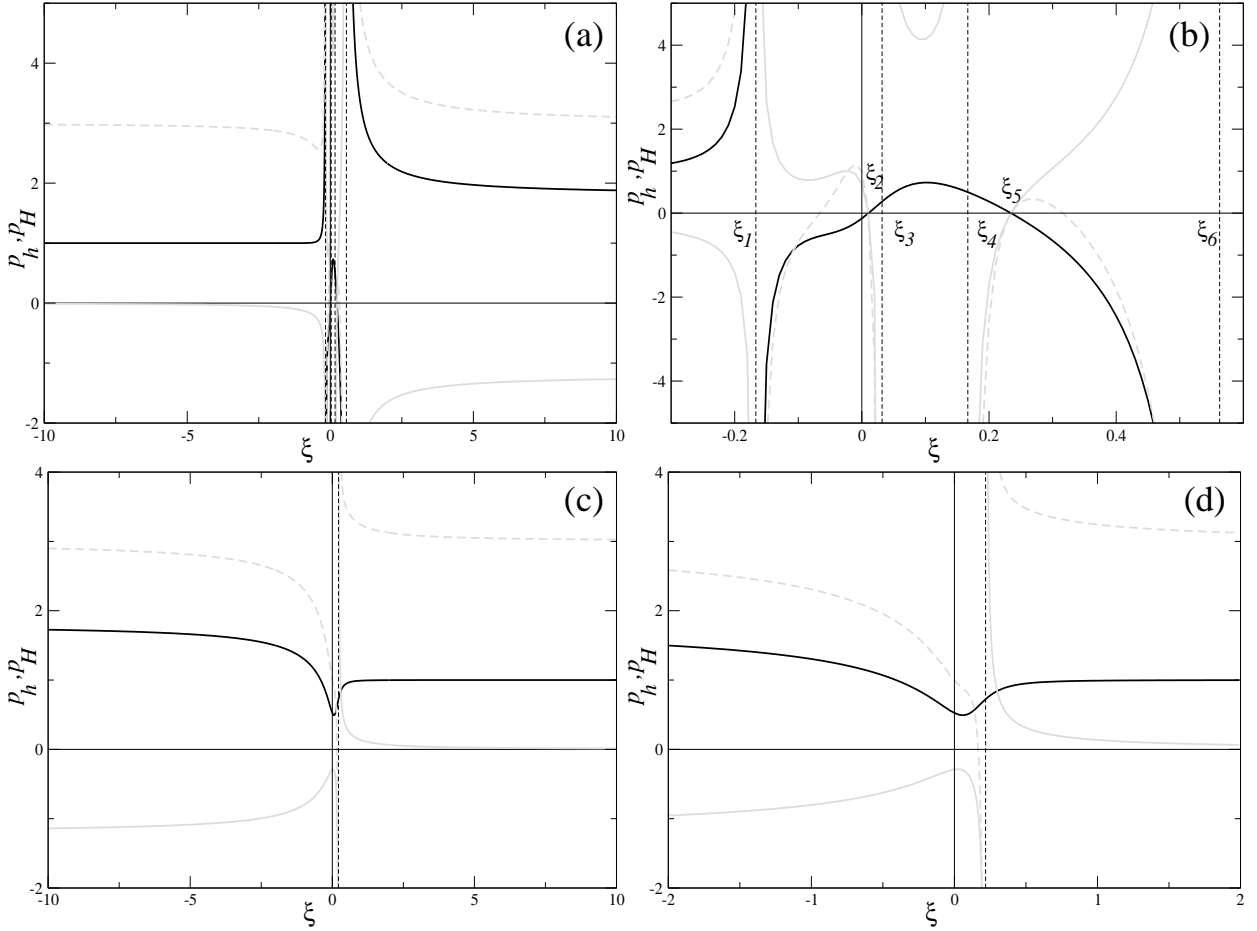}
\caption{Dynamics of Kasner exponents in $D=2$ vacuum model. In (a) and (b) panels we presented large-scale structure (a) and fine structure (b) for $h_+$ branch while in (c) and (d) -- the same but for
$h_-$ branch -- large-scale in (c) and fine structure in (d). Black line corresponds to $p_H$, solid grey -- to $p_h$ and dashed grey -- to the expansion rate $3p_H + 2p_h$ (see text for details).}\label{D2p}
\end{figure}

We presented individual Kasner exponents as well as the expansion rate in Fig.~\ref{D2p}. In (a) and (b) panels we presented the dynamics for $h_+$ branch -- large-scale structure in (a) and fine structure
in (b) panel. In (c) and (d) panels we have the same but for $h_-$ branch -- large-scale structure in (c) and fine structure in (d). Black lines correspond to $p_H$, solid grey -- to $p_h$ and dashed grey
depict the expansion rate $3p_H + 2p_h$.

From (a) and (c) panels of Fig.~\ref{D2p} we can immediately confirm Gauss-Bonnet Kasner regime as high-energy asymptote in all cases -- and confirm corresponding Kasner exponents from Table~\ref{D.21}.
Panel (b) of Fig.~\ref{D2p} corresponds to $h_+$ branch -- $\xi < 0$ part depicts $\alpha < 0$ case while $\xi > 0$ corresponds to $\alpha > 0$. Now we can compare $\xi < 0$ part of Fig.~\ref{D2p}(b) with
Fig.~\ref{D2v}(e) -- they both represent the same dynamical behavior but in different coordinates. One can see the isotropic exponential solution at $\xi_1 = -1/6$, but cannot see nonstandard
singularity at $\xi = -1/36$. It is present in $\dot H$, $\dot h$ analysis but absent in $p_H$, $p_h$ due to cancellations -- this illustrates the fact that for full analysis we cannot rely on any metric
{\it ansatz} and should perform it in maximally general case.

Continuing with comparisons, we compare $\xi > 0$ part of  Fig.~\ref{D2p}(b) with Fig.~\ref{D2v}(c) -- $\xi_2 \approx 0.01046$ from $Q^+ = 0$ (\ref{D2_Qplus}) is zero-point for all Kasner exponents as it
corresponds to nonstandard singularity $H_1$ from Fig.~\ref{D2v}(c); two vertical asymptotes $\xi_3 = 1/8 - \sqrt{5}/24 \approx 0.03183$ and $\xi_4 = 1/6$ (from roots of $P_2^+$) correspond to $\dot h = 0$
between nonstandard singularities $H_1$ and $H_2$ from Fig.~\ref{D2v}(c); zero-point $\xi_5 \approx 0.23449$ from $Q^+ = 0$ (\ref{D2_Qplus}) correspond to nonstandard singularity $H_2$ from
Fig.~\ref{D2v}(c) and finally asymptote $\xi_6 \approx 0.56276$ (from roots of $P_2^+$) correspond to the isotropic exponential solution at $H_3$ of Fig.~\ref{D2v}(c).

The $h_-$ branch has less abundant dynamics; one can simply map $\xi < 0$ part of Fig.~\ref{D2p}(d) into Fig.~\ref{D2v}(f) while its $\xi > 0$ part also could be seen as Fig.~\ref{D2v}(d) remembering that
$\dot h = 0$ point in Fig.~\ref{D2v}(d) occurs at $\xi = 1/8 + \sqrt{5}/24 \approx 0.21817$ which comes from $P_2^-$ roots.

With this we mapped the dynamics in $\{\dot H,\,\dot h\}$ coordinates with $\{p_H,\,p_h\}$ one. We saw that this mapping is not entirely equivalent as it could ``lose'' or ``create'' singularities.



We summarize our findings for $D=2$ regimes in Table \ref{D.2}. In addition to columns we had in Table \ref{D.1}, we added ``Branch'' column since starting from $D=2$ case we have several branches
for $h$ solutions from constraint -- e.g. two in $D=2$ case (\ref{D2_hvac}). Also we denoted exponential regime with separate expansions rates from $h_+$, $\alpha > 0$ case as $E_{3+2}$ to
distinguish it from isotropic exponential solution $E_{iso}$.

\begin{table}[h]
\begin{center}
\caption{Summary of $D=2$ regimes.}
\label{D.2}
  \begin{tabular}{|c|c|c|c|c|}
    \hline
     $\alpha$ &Branch   & \multicolumn{2}{c|}{Additional conditions} & Regimes  \\
    \hline
 \multirow{5}{*}{$\alpha > 0$} & \multirow{4}{*}{$h_+$} &  \multicolumn{2}{c|}{$H < H_1 = \sqrt{\frac{\xi_1}{\alpha}}$ from (\ref{D2_Qplus}) } & $K_1 \to nS$  \\  \cline{3-5}
 & &  \multicolumn{2}{c|}{$\sqrt{\frac{\xi_2}{\alpha}} = H_2 > H > H_1 = \sqrt{\frac{\xi_1}{\alpha}}$ from (\ref{D2_Qplus}) } & $nS \to nS$  \\   \cline{3-5}
 & &  \multicolumn{2}{c|}{$\sqrt{\frac{\xi_0}{\alpha}} = H_3 > H > H_2 = \sqrt{\frac{\xi_2}{\alpha}}$ from (\ref{D2_roots}) and (\ref{D2_Qplus}) } & $nS \to E_{3+ 2}$  \\  \cline{3-5}
 & &  \multicolumn{2}{c|}{ $ H > H_3 = \sqrt{\frac{\xi_0}{\alpha}}$ from (\ref{D2_roots}) } & $K_3 \to E_{3+ 2}$  \\ \cline{2-2}  \cline{3-5}
 & \multirow{2}{*}{$h_-$} &  \multicolumn{2}{c|}{\multirow{2}{*}{no}} & \multirow{2}{*}{$K_3 \to K_1$}  \\  \cline{1-1}
 \multirow{4}{*}{$\alpha < 0$} &  &\multicolumn{2}{c|}{} &  \\  \cline{2-2}  \cline{3-5}
 & \multirow{3}{*}{$h_+$} &  \multicolumn{2}{c|}{$H < H_1 = \frac{1}{6\sqrt{-\alpha}}$} & $K_1 \to nS$  \\ \cline{3-5}
 & &  \multicolumn{2}{c|}{$\frac{1}{\sqrt{-6\alpha}} = H_2 > H > H_1 = \frac{1}{6\sqrt{-\alpha}}$} & $nS \to E_{iso}$  \\ \cline{3-5}
 & &  \multicolumn{2}{c|}{$H > H_2 = \frac{1}{\sqrt{-6\alpha}}$} & $K_3 \to E_{iso}$  \\
\hline
  \end{tabular}
\end{center}
\end{table}

To conclude, in $D=2$ vacuum regime there are total 8 different regimes but only three of them are nonsingular -- $K_3 \to K_1$, $K_3 \to E_{3+2}$ and $K_3 \to E_{iso}$. The first of them is natural
and the only regime for
$h_-$ branch while the remaining occur for $h_+$ branch for either $\alpha > 0$ (anisotropic) and $\alpha < 0$ (isotropic) and with different bounds on $H$ (see Table \ref{D.2}).
Also of these three only $K_3 \to K_1$ and $K_3 \to E_{3+2}$ could be viable -- $K_3 \to E_{iso}$ one expects
isotropisation of all spatial dimensions and that is not what we observe nowadays. Let us also note that both exponential regimes appear within their stability ranges found in~\cite{my15}.

\section{$D=3$ case}\label{S3}

In this case the equations of motion take form ($H$-equation, $h$-equation and constraint correspondingly):


\begin{equation}
\begin{array}{l}
4\dot H + 6H^2 + 6\dot h + 12h^2 + 12Hh + 8\alpha \( 6\dot Hh (H + h) + 3\dot h (H^2 + h^2 + 4Hh) +18H^2h^2 + \right.\\ \left. +18Hh^3 + 3h^4 + 6H^3h\)  = 0,
\end{array} \label{D3_H}
\end{equation}

\begin{equation}
\begin{array}{l}
6\dot H + 12H^2 + 4\dot h + 6h^2 + 12Hh + 8\alpha \( 3\dot H (H^2 + 4Hh + h^2) + 6\dot h H (H+h) + 6Hh^3 + \right. \\ \left. + 18H^2h^2 + 18H^3h + 3H^4  \) = 0,
\end{array} \label{D3_h}
\end{equation}

\begin{equation}
\begin{array}{l}
6 H^2 + 18Hh + 6h^2 + 24\alpha (3H^3h + 9H^2h^2 + 3Hh^3 ) = 0.
\end{array} \label{D3_con}
\end{equation}


Solving the constraint equation (\ref{D3_con}) for $h$, one gets

\begin{equation}
\begin{array}{l}
h_1 = - \dac{1}{12\alpha H}; \; h_{2,3} = \( -\dac{3}{2}\pm\dac{\sqrt{5}}{2}\)H,
\end{array} \label{D3_h_vac}
\end{equation}

\noindent with ``+'' sign corresponds to $h_2$ and ``--'' to $h_3$. One can see that $h_{2,\,3}$ always have opposite sign from $H$ while $h_1$ has opposite sign for $\alpha > 0$ and the same
for $\alpha < 0$.

Now we can solve (\ref{D3_H})--(\ref{D3_h}) with respect to $\dot H$ and $\dot h$ and substitute branches obtained (\ref{D3_h_vac}) to get expressions for $\dot H(H^2)$ and $\dot h(H^2)$:

\begin{equation}
\begin{array}{l}
\dot H_1 = - \dac{1}{12\alpha} \times \dac{1728\xi^3 + 1}{144\xi^2 + 12\xi + 1},~\dot h_1 = - \dac{1}{144\alpha\xi}\times \dac{1728\xi^3 + 1}{144\xi^2 + 12\xi + 1}, \\ \\
\dot H_{2,\,3} = \dac{3\xi}{2\alpha} \dac{P_1^\mp}{Q^\mp},~\dot h_{2,\,3} = - \dac{3\xi}{2\alpha} \dac{P_2^\mp}{Q^\mp} ~~\mbox{with}~~\xi = \alpha H^2~~\mbox{and} \\
P_1^\mp = 240\sqrt{5}\xi^2 \mp 528\xi^2 - 32\sqrt{5}\xi \pm 64\xi + \sqrt{5} \mp 1, \\ P_2^\mp = 312\sqrt{5}\xi^2 \mp 696\xi^2 - 40\sqrt{5}\xi \pm 88\xi + \sqrt{5} \mp 2, \\
Q^\pm = 216\sqrt{5}\xi^2 \mp 504\xi^2 - 12\sqrt{5}\xi \pm 36\xi \mp 1;
\end{array} \label{D3_dH_vac}
\end{equation}

\noindent signs follow (\ref{D3_h_vac}) notation: upper corresponds to ``2'' subscript while lower -- to ``3''.

\begin{figure}
\includegraphics[width=1.0\textwidth, angle=0]{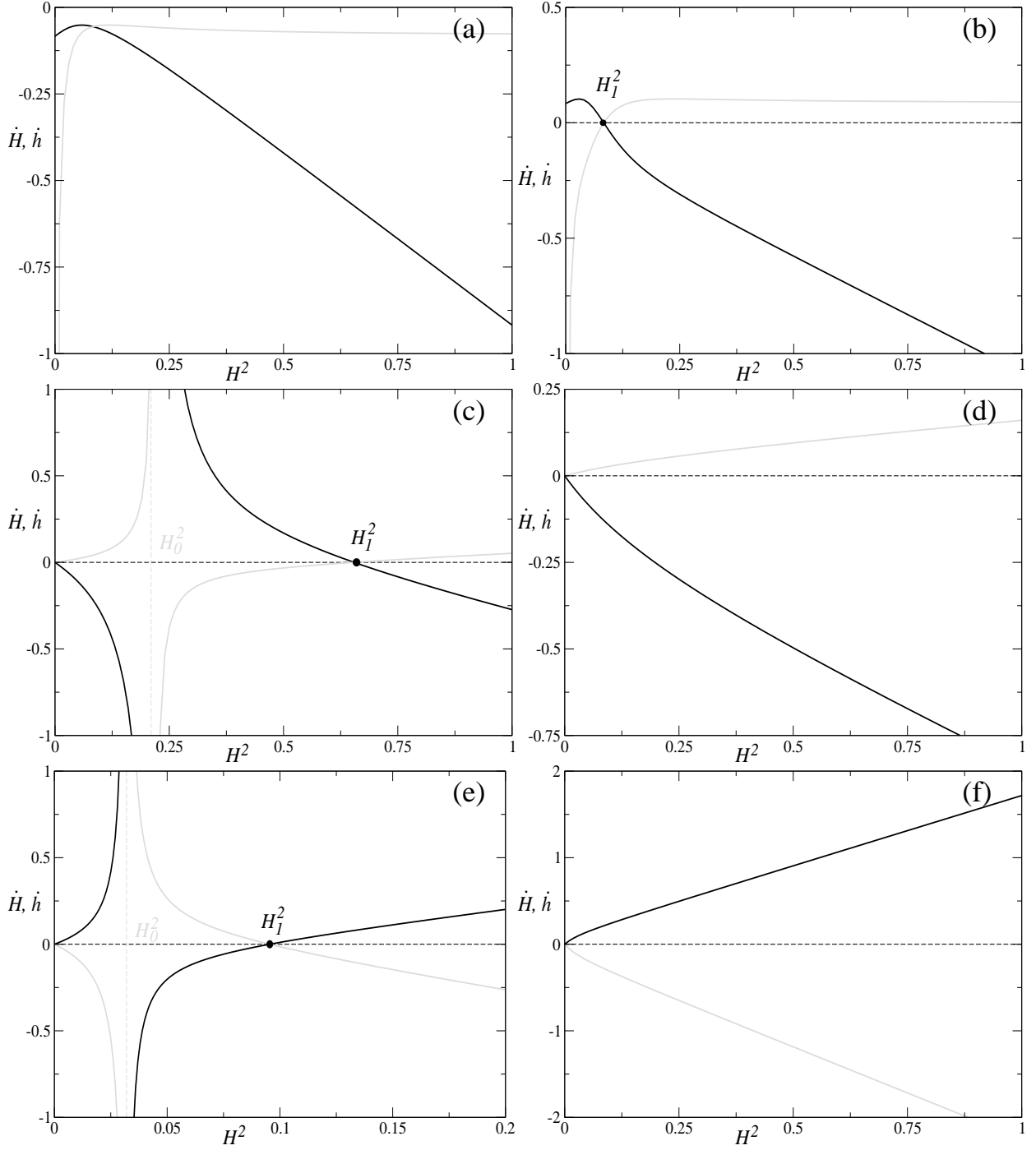}
\caption{The plot of $\dot H$ (black curve) and $\dot h$ (grey curve) vs $H^2$ for $D=3$ vacuum model. In (a) panel we plotted $h_1$ branch with $\alpha > 0$, in (b) panel -- $h_1$ branch with
$\alpha < 0$, (c) -- $h_2$, $\alpha > 0$, (d) -- $h_2$, $\alpha < 0$, (e) -- $h_3$, $\alpha > 0$, (f) -- $h_3$, $\alpha < 0$ (see text for details).}\label{D3v}
\end{figure}

Before analyzing $\dot H$ and $\dot h$ vs $H^2$ in a plot it is useful to find their zeros and asymptotes -- roots of $P_{1,\,2}^\pm$ and $Q^\pm$ respectively. Here they are:

\begin{equation}
\begin{array}{l}
P_{1,\,2}^+: \xi = \xi_4 = \dac{\sqrt{5} - 2}{2(5\sqrt{5} - 11)} \approx 0.65451;~ \xi = \xi_3 = \dac{\sqrt{5} - 2}{6(5\sqrt{5} - 11)} \approx 0.21817; \\
P_{1,\,2}^-: \xi = \xi_2 = \dac{\sqrt{5} + 2}{2(5\sqrt{5} + 11)} \approx 0.09549;~ \xi = \xi_1 = \dac{\sqrt{5} + 2}{6(5\sqrt{5} + 11)} \approx 0.03183; \\
Q^+: \xi = \xi_3;~Q^-: \xi=\xi_1;
\end{array} \label{D3_roots}
\end{equation}

\noindent and numbering of $\xi$ is arranged for growing values of $\xi$. Additionally, $\dot H_1 = 0$ as well as $\dot h_1 = 0$ give us $\xi = -1/12$ root.

Now we can plot $\dot H$ and $\dot h$ vs $H^2$ -- see Fig. \ref{D3v}. There in (a) panel we presented $\dot H(H^2)$ and $\dot h(H^2)$ curves for
$h=h_1$ branch and $\alpha > 0$ while in (b) panel -- the same branch but with $\alpha < 0$ choice. Similar structure have two remaining rows -- (c) panel corresponds to $h=h_2$ branch with $\alpha > 0$
while (d) panel -- $h=h_2$ with $\alpha < 0$. Finally, (e) panel depicts graphs for $h=h_3$ branch with $\alpha > 0$ and (f) -- $h=h_3$ with $\alpha < 0$.

Let us have a closer look on the panels. In (a) panel ($h=h_1$, $\alpha > 0$) both $\dot H$ and $\dot h$ are always negative; $\dot h$ is also singular at $H=0$. Additional studies of $p_h$ and $p_H$
(see below) reveal that $H=0$ is Gauss-Bonnet Kasner singularity.
In (b) panel ($h=h_1$, $\alpha < 0$) we have stable point at $H_1^2 = - \dac{1}{12\alpha}$; this point corresponds to isotropic exponential expansion:
$H,\,h \to - \sqrt{\dac{1}{12\alpha}}$. This regime is reached from both sides -- $H < H_1$ and $H > H_1$ and in both cases the past regime is singular high-energy (Gauss-Bonnet) Kasner regime $K_3$.
In (c) panel ($h=h_2$, $\alpha > 0$), we have $H,\,h\to 0$ solution with $\dac{h}{H} = \( -\dac{3}{2}+\dac{\sqrt{5}}{2}\)$ (see (\ref{D3_h_vac})) for
$H^2 < H_0^2 = \dac{\sqrt{5}+3}{24\alpha}$; this appears to be GR Kasner solution $K_1$.
For $H^2 > H_0^2$ we have stable point $H_1^2 = \dac{\sqrt{5}+3}{8\alpha}$ with exponential solution, but past asymptotes are different -- for $H_0 < H < H_1$ it is nonstandard singularity while for
$H > H_2$ it is GB Kasner regime $K_3$. Exponential solution corresponds to $H^2 = \dac{\sqrt{5}+3}{8\alpha}$, and from (\ref{D3_h_vac}) we know that $h/H = -3/2 + \sqrt{5}/2$, so it is anisotropic exponential
solution.
In panel (d) (($h=h_2$, $\alpha < 0$)), at $H\to 0$, we have the same
behavior as first regime in (c) panel: $H,\,h\to 0$ with $\dac{h}{H} = \( -\dac{3}{2}+\dac{\sqrt{5}}{2}\)$, but the past asymptote is different -- now it is GB Kasner regime $K_3$.
Last two cases correspond to $h=h_3$: $\alpha > 0$ in (e) panel and $\alpha < 0$ in (f). One can see that they are in a sense ``reverse'' of $h_2$ regimes --
first regime in (e) panel is singular: for $H^2 < H_1^2 = \dac{3-\sqrt{5}}{8\alpha}$ we have attractor $H^2 = H_0^2 = \dac{3-\sqrt{5}}{24\alpha}$ where both $\dot H$ and $\dot h$ diverge, so it is
$K_1 \to nS$, which is directly opposite to what we saw in (c) panel. Two remaining regimes have exponential solution as past asymptotes (opposite to future asymptotes in (c) panel) and either
non-standard singularity for $H_0 < H < H_1$ or GB Kasner for $H > H_1$. One can clearly see the difference between (c) and (e) panels -- future and past asymptotes interchange. The same is in
(f) panel -- there we have GR Kasner as past and GB Kasner as future asymptotes, exactly opposite to (d) panel.

Finally, similarly to the previous sections, we rewrite equations of motion in terms of Kasner exponents $p_H$ and $p_h$ to see Kasner asymptotes. Following the definition $p = - H^2/\dot H$ with use of
(\ref{D3_h_vac}) and (\ref{D3_dH_vac}) we can write down exact expressions for $p_H$ and $p_h$:

\begin{equation}
\begin{array}{l}
p_{H, 1} = \dac{12\xi (144\xi^2 + 12\xi + 1)}{1728\xi^3 + 1},~p_{h, 1} = \dac{144\xi^2 + 12\xi + 1}{1728\xi^3 + 1}; \\
p_{H, (2,3)} = p_H^\pm = -\dac{H^2}{\dot H_\pm} = - \dac{2 Q^\pm}{3 P_1^\pm},~p_{h, (2,3)} = p_h^\pm = -\dac{h_\pm^2}{\dot h_\pm} = \dac{\left(3\pm \dac{\sqrt{5}}{2}\right)^2}{3} \dac{Q^\pm}{P_2^\pm}.
\end{array} \label{D3_pHph}
\end{equation}

By taking appropriate limits from (\ref{D3_pHph}) and confirming them with the results of direct computations of $\lim \dot H/H^2$ and $\lim \dot h/h^2$ (similar to the previous sections) we can
find power-law behavior in the limiting cases $H\to 0$ and $H\to\infty$; the results are presented in Table~\ref{D.31}. Let us note that unlike previous cases, for $D=3$ limits for $\alpha >0$ and
$\alpha < 0$ coincide.

\begin{table}[h]
\begin{center}
\caption{Power-law behavior in $D=3$}
\label{D.31}
  \begin{tabular}{|c|c|c|c|c|}
    \hline
     $h\pm$    & $\lim H$ & $p_H$ & $p_h$ & $\sum p_i$   \\
    \hline
\multirow{2}{*}{$h_1$}  & 0        & $0$ & $1$ & 3 \\  \cline{2-5}
                        & $\infty$ & $1$             & $0$                  & 3 \\  \cline{1-5}
\multirow{2}{*}{$h_+$}  & 0        & $\frac{2}{3\sqrt{5}-3}$ & $\frac{3\sqrt{5}-7}{6\sqrt{5}-12}$ & 1 \\  \cline{2-5}
			& $\infty$ & $\frac{7-3\sqrt{5}}{5\sqrt{5}-11}$             & $\frac{21\sqrt{5}-47}{13\sqrt{5}-29}$                  & 3 \\  \cline{1-5}
\multirow{2}{*}{$h_-$}  & 0        & $-\frac{2}{3\sqrt{5}+3}$ & $\frac{3\sqrt{5}+7}{6\sqrt{5}+12}$ & 1 \\  \cline{2-5}
			& $\infty$ & $-\frac{7+3\sqrt{5}}{5\sqrt{5}+11}$                        & $\frac{21\sqrt{5}+47}{13\sqrt{5}+29}$                               & 3 \\
\hline
  \end{tabular}
\end{center}
\end{table}

\begin{figure}
\includegraphics[width=1.0\textwidth, angle=0]{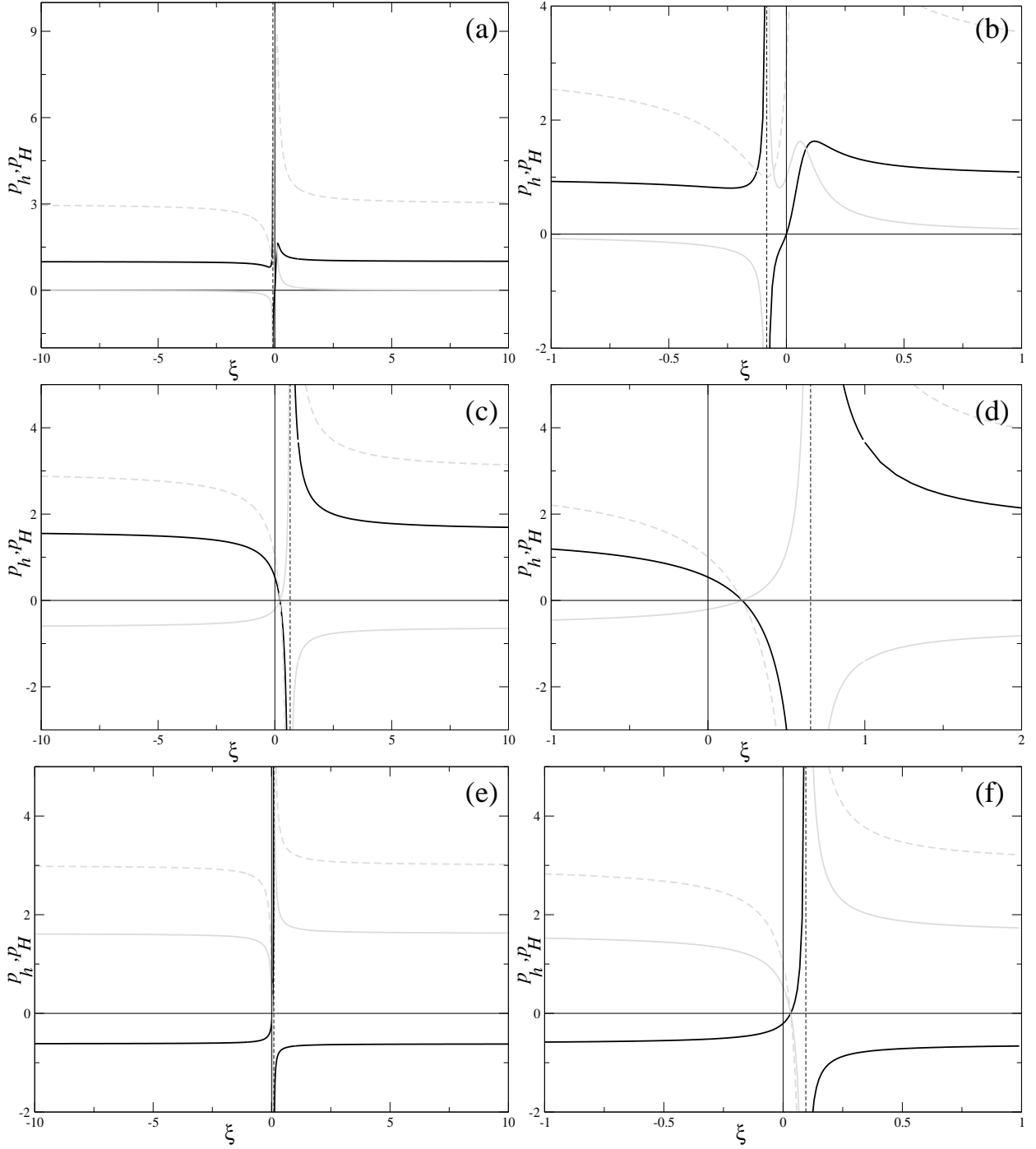}
\caption{Dynamics of Kasner exponents in $D=3$ vacuum model. In (a) and (b) panels we presented large-scale structure (a) and fine structure (b) for $h_1$ branch, in (c) and (d) -- large-scale structure (c)
and fine structure (d) for $h_+$ branch and in (e) and (f) panels we presented large-scale structure (e) and fine structure (f) for $h_-$ branch (see text for details).}\label{D3_p}
\end{figure}

Now we can plot $p_H$ and $p_h$ versus $\xi$ for all three branches of $h$; the corresponding plots are presented in Fig.~\ref{D3_p}. The first row ((a) and (b) panels) corresponds to first $h_1$ branch
of Eq.~(\ref{D3_h_vac}), the second ((c) and (d) panels) -- to $h_+$ and the last ((e) and (f) panels) to $h_-$. The first column ((a), (c) and (e) panels) gives large-scale behavior while the second
((b), (d) and (f) panels) -- fine-scale in the vicinity of $\xi = 0$. From first column one can verify our limits for $H\to\infty$ from
Table~\ref{D.31} while from the second column -- our limits for $H\to 0$. Also one can see that the limits for $H\to\infty$ coincide for $\alpha > 0$ and $\alpha < 0$.

Similar to the previous sections let us make mappings between the dynamics in $\{\dot H,\,\dot h\}$ and $\{p_H,\,p_h\}$ coordinates. The first row corresponds to $h_1$ branch and so to (a) and (b) panels of
Fig.~\ref{D3v}. For $\alpha > 0$ (Fig.~\ref{D3v}(a)) which corresponds to $\xi > 0$ we have singular $K_3 \to K_3$ transition and that is what we see in Fig.~\ref{D3_p}(b) -- one can see that at $\xi = 0$
we have $\sum p = 3$, which makes it GB Kasner. For $\alpha < 0$ (Fig.~\ref{D3v}(b)) we have isotropic exponential solution with GB Kasner from both sides -- and that is exactly what we see from $\xi < 0$
part of Fig.~\ref{D3_p}(b). But there is an interesting feature -- at exponential solution Kasner exponents diverge -- and we see that both $p_H$ and $p_h$ are divergent -- but their sum $\sum p$ is not.
It is an artifact caused by the fact that the number of dimensions in both manifolds is the same -- previously individual exponents did not cancel each other in this way at exponential solutions.

The second and third rows corresponds to $h_\pm$ branches. While describing them in $\{\dot H,\,\dot h\}$ coordinates we noted that they are ``reverted'' in a way. The same is true for the description in
$\{p_H,\,p_h\}$ coordinates -- one cannot miss similarity between Figs.~\ref{D3_p}(c, d) and Figs.~\ref{D3_p}(e, f). Both Figs.~\ref{D3_p}(d, f) in $\xi < 0$ have transition $K_1 \to K_3$ -- the
same we see in Figs.~\ref{D3v}(d, f) ($h_\pm$, $\alpha < 0$) with the difference that in Fig.~\ref{D3v}(f), which is ($h_-$, $\alpha < 0$), it is reverted with respect to time. In $\{p_H,\,p_h\}$ coordinates,
the difference is in interchanging between $p_H$ and $p_h$. Similar effects we observe in $\xi > 0$ part -- nonstandard singularities in Figs.~\ref{D3v}(c, e) correspond to $p_H = p_h \equiv 0$ in
\mbox{Figs.~\ref{D3_p}(d, f)} and the exponential solutions in Figs.~\ref{D3v}(c, e) are mapped into vertical asymptotes in \mbox{Figs.~\ref{D3_p}(d, f)}. Similar to the previously described case, the difference
between Figs.~\ref{D3v}(c, e) is ``time reversal'' which is represented as interchanging between $p_H$ and $p_h$ in $\{p_H,\,p_h\}$ coordinates in Figs.~\ref{D3_p}(d, f).

We summarize our findings for $D=3$ regimes in Table \ref{D.3}. The denotations are similar to the previous case -- Table~\ref{D.2}.

\begin{table}[h]
\begin{center}
\caption{Summary of $D=3$ regimes.}
\label{D.3}
  \begin{tabular}{|c|c|c|c|c|}
    \hline
     Branch & $\alpha$  & \multicolumn{2}{c|}{Additional conditions} & Regimes  \\
    \hline
 \multirow{2}{*}{$h_1$} & $\alpha > 0$ &  \multicolumn{2}{c|}{\multirow{2}{*}{no}} & $K_3 \to K_3$  \\ \cline{2-2} \cline{5-5}
 & $\alpha < 0$  &  \multicolumn{2}{c|}{} & $K_3 \to E_{iso}$ (both branches)  \\ \cline{1-5}
 \multirow{4}{*}{$h_2$} & \multirow{3}{*}{$\alpha > 0$}  & \multicolumn{2}{c|}{$ H < \sqrt{\dac{\xi_3}{\alpha}}$ from (\ref{D3_roots})} & $nS \to K_1$  \\  \cline{3-5}
 &  & \multicolumn{2}{c|}{$\sqrt{\dac{\xi_4}{\alpha}} > H > \sqrt{\dac{\xi_3}{\alpha}}$ from (\ref{D3_roots})} & $nS \to E_{3+ 3}$  \\ \cline{3-5}
 &  & \multicolumn{2}{c|}{$H > \sqrt{\dac{\xi_4}{\alpha}}$ from (\ref{D3_roots})} & $K_3 \to E_{3+ 3}$ \\ \cline{2-5}
 & $\alpha < 0$  & \multicolumn{2}{c|}{no} & $K_3 \to K_1$  \\ \cline{1-5}
 \multirow{4}{*}{$h_3$} & \multirow{3}{*}{$\alpha > 0$}   & \multicolumn{2}{c|}{$H < \sqrt{\dac{\xi_1}{\alpha}}$ from (\ref{D3_roots})} & $K_1 \to nS$  \\ \cline{3-5}
 &  & \multicolumn{2}{c|}{$\sqrt{\dac{\xi_2}{\alpha}} > H > \sqrt{\dac{\xi_1}{\alpha}}$ from (\ref{D3_roots})} & $E_{3+ 3} \to nS$  \\  \cline{3-5}
 &  & \multicolumn{2}{c|}{$H > \sqrt{\dac{\xi_2}{\alpha}}$ from (\ref{D3_roots})} & $E_{3+ 3} \to K_3$  \\  \cline{2-5}
 & $\alpha < 0$   & \multicolumn{2}{c|}{no} & $K_1 \to K_3$  \\
\hline
  \end{tabular}
\end{center}
\end{table}

To conclude, in $D=3$ vacuum model we have 10 different regimes, but some of them have more then one branch, like $h_1$ $\alpha < 0$ case (see Fig.~\ref{D3v}(b)), when both regimes are $K_3 \to E_{iso}$
but in one of them $H$ is increasing and in another it is decreasing. But on the other hand both subspaces are three-dimensional so we cannot discriminate them. Another interesting point is that
of three branches of solutions one ($h_1$) is ``independent'' while two remaining are linked through some sort of ``reversal''. Exactly, all regimes in these two branches coincide up to ``time reversal'' --
if in $h_+$ we have, say, $nS \to K_1$ transition, in $h_-$ we have $K_1 \to nS$. We link this feature with the structure of the solutions -- the constraint equation (\ref{con2_gen}) is cubic with respect
to $H$ and so the structure of its solutions affect solutions of the entire system.

Of ten different regimes only three are nonsingular -- $K_3 \to E_{iso}$ with $\alpha < 0$ on $h_1$ branch, \mbox{$K_3 \to E_{3+3}$} with $\alpha > 0$ and $K_3 \to K_1$ with $\alpha < 0$ -- both from $h_+$ branch. 
Of these three regimes isotropisation is not viable -- we do observe discrimination between three and extra dimensions, so only two regimes (and both of them are from $h_+$ branch) remain.
As we described, $h_-$ branch is reverse of $h_+$, so there are no viable regimes on $h_-$ branch as well.

\section{General $D\geqslant 4$ case}

\begin{figure}
\includegraphics[width=1.0\textwidth, angle=0]{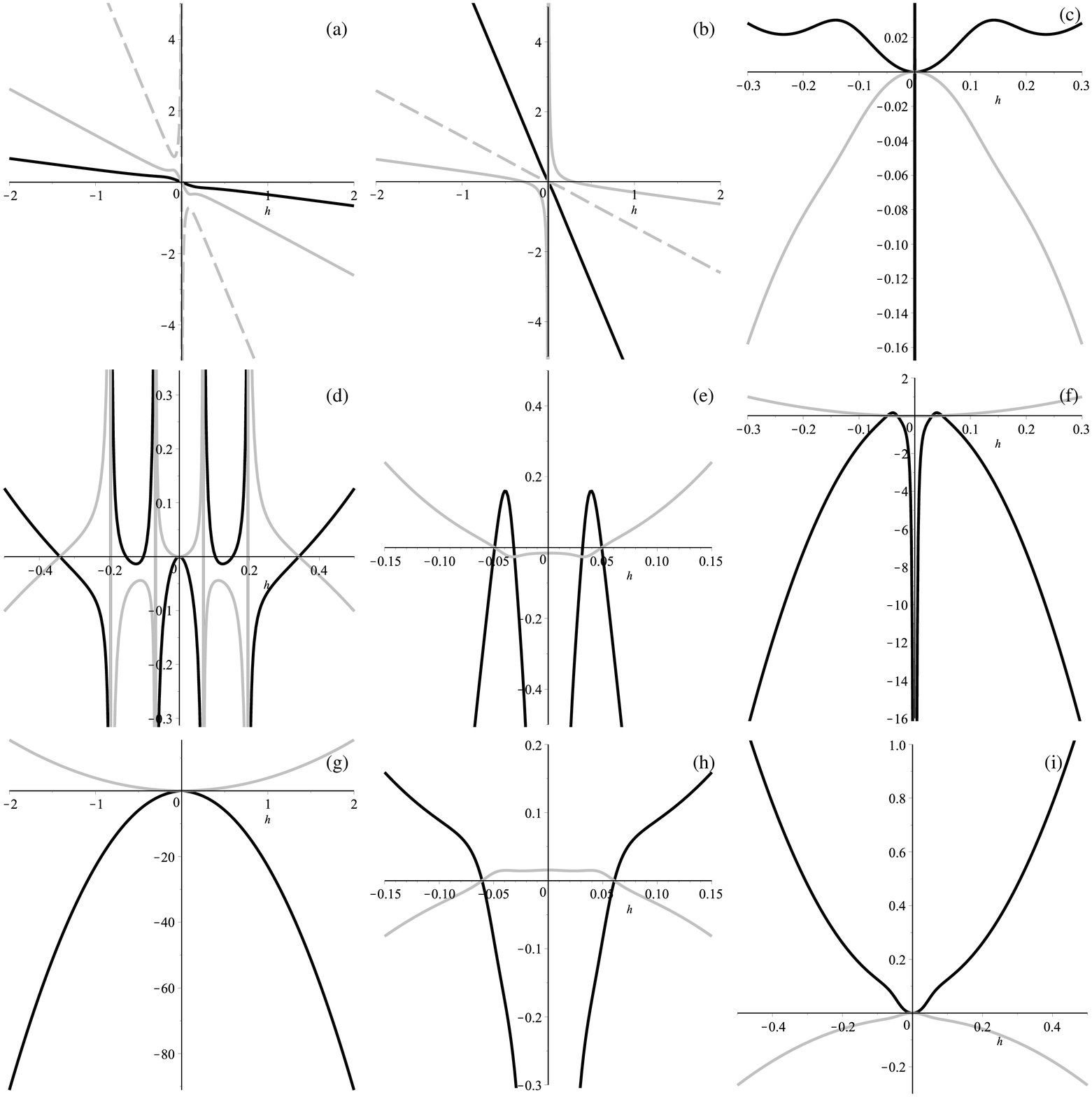}
\caption{Typical dynamics for $D\geqslant 4$ case. In (a) and (b) panel we presented $H(h)$ curves -- $\alpha > 0$ in (a) and $\alpha < 0$ -- in (b). Three different branches ($H_1$, $H_2$ and $H_3$) are
presented in three different linestyles -- black, solid grey and dashed grey. In the remaining panels we presented $\dot H$ (in black) and $\dot h$ (in grey) curves for $\alpha > 0$ in (c)--(f) and for
$\alpha < 0$ in (g)--(i) panels. Panel (c) corresponds to $H_1$, panel (d) -- to $H_2$, panels (e) and (f) -- to $H_3$ -- fine structure in the vicinity of $h=0$ in (e) and large-scale structure in (f).
For $\alpha < 0$ (g) panel represents $H_1$, (h) -- $H_2$ and (i) -- $H_3$.
(see text for details).}\label{D4_1}
\end{figure}

In the general case, we use general equations (\ref{H_gen})--(\ref{con2_gen}), but unlike previous cases we solve (\ref{con2_gen}) with respect to $H$ instead of $h$. The reason for it is that now
(\ref{con2_gen}) is cubic with respect to $H$ but quartic with respect to $h$, and now it is simpler to solve it with respect to $H$. The general form of the solution is complicated, but we can plot resulting
$H(h)$ curves. Typical $H(h)$ curves for $D\geqslant 4$ case are given in Figs.~\ref{D4_1} (a, b) -- $\alpha > 0$ on (a) and $\alpha < 0$ on (b). In there we put three branches ($H_1$, $H_2$ and
$H_3$ -- as three solutions of cubic equations) with different colors and linestyles -- black, solid grey and dashed grey.
One can see that the situation resemble $D=3$
rather then $D=1$ or $D=2$. Exact curves in Fig.~\ref{D4_1}(a, b) correspond to $D=6$, but for any other $D\geqslant 4$ typical behavior is the same, the difference lies only in the inclination of
asymptotic $h \to \pm \infty$ behavior.

The procedure is similar to the previous cases -- we solve (\ref{H_gen}) and (\ref{h_gen}) with respect to $\dot h$ and $\dot H$, but unlike previous cases we substitute not $h(H)$, but $H(h)$ now. The
difference is the following -- in previous cases $H$ was the ``dynamical variable'' and so $\dot H(H)$ leads the evolution while $\dot h(H)$ followed. Now our ``dynamical variable'' is $h$ and so
$\dot h(h)$ leads while $\dot H(h)$ follows. Similar to the $H(h)$ functions, functional form of $\dot h$ and $\dot H$ is too complicated to write them down, so we substitute $H(h)$ into them and plot the
resulting curves $\dot h(h)$ and $\dot H(h)$ in Fig.~\ref{D4_1}(c)--(i).

Panels (c)--(f) correspond to $\alpha > 0$ while panels (g)--(i) -- to $\alpha < 0$. Black curves correspond to $\dot H(h)$ and grey -- to $\dot h(h)$. Now let us have a closer look on the panels.
Remember, from Fig.~\ref{D4_1}(a, b) we found that for $H > 0$ one needs $h < 0$ in most cases, so we take closer look on $h < 0$ part. In (c)
panel, which corresponds to $H_1$, $\alpha > 0$, we can clearly see $K_1 \to K_3$ transition (we prove it with $p_H$ and $p_h$ graph below). In (d) panel, which correspond to $H_2$ with $\alpha > 0$, the
situation is more complicated -- we have (with increase of absolute value for $h$) $nS \to K_1$, then $nS \to nS$, $E \to nS$ and finally $E \to K_3$. One can clearly see that the exponential solution is unstable for $H > 0$,
but is stable for $H < 0$ -- see $h > 0$ part of Fig.~\ref{D4_1}(d); from~\cite{my15} we know that, say, for $(4+3)$ splitting (and it falls within $D\geqslant 4$ case) there are three exponential
solutions -- isotropic which is stable for $H > 0$ and two anisotropic solutions with $(4+3)$ splitting -- one of them is stable for $H > 0$ and the other is unstable; we can assume that the general
$(D+3)$ splitting has the same property -- one stable (for $H > 0$) isotropic solution and two anisotropic solution with one of the stable for $H > 0$ and the other -- for $H < 0$. This way, the
exponential solution found is anisotropic; further on, our numerical investigation proves it.

Next two panels -- (e) and (f) -- correspond to $H_3$ with $\alpha > 0$. There are two regimes but they seem to be the same -- $K_3 \to E$. This exponential solution is also anisotropic and our numerical
investigation proves it. One could mistake regime in $h\to 0$ with $K_1$ but our investigation with $p_H$ and $p_h$ (see below) shows that it is $K_3$. The remaining panels correspond to $\alpha < 0$: in (g) panel
($H_1$) we can clearly see $K_3 \to K_1$ regime; in (h) panel ($H_2$) there are two regimes with unstable isotropic exponential solution and both of them are $E_{iso} \to K_3$. In the $h\to 0$ limit one could
assume that the regime is $K_1$ but analysis in ($p_H$, $p_h$) coordinates reveals that it is $K_3$. The mentioned unstable exponential solution turn into stable at $h>0$ and the ``correct'' solution is namely $h>0$ --
indeed, from solid grey curve in Fig.~\ref{D4_1}(b) (which corresponds to the $H_2$ branch which we are dealing with right now) one can see that to have $H>0$ in the vicinity of $h=0$ we need $h>0$. In this way, unstable isotropic solution is replaced with the stable one and the desciption follows the general scheme~\cite{my15}.
Finally, in (i) panel, which corresponds to $H_3$, $\alpha < 0$, we can see $K_1 \to K_3$ -- similar to (c) panel.

\begin{figure}[h!]
\includegraphics[width=1.0\textwidth, angle=0]{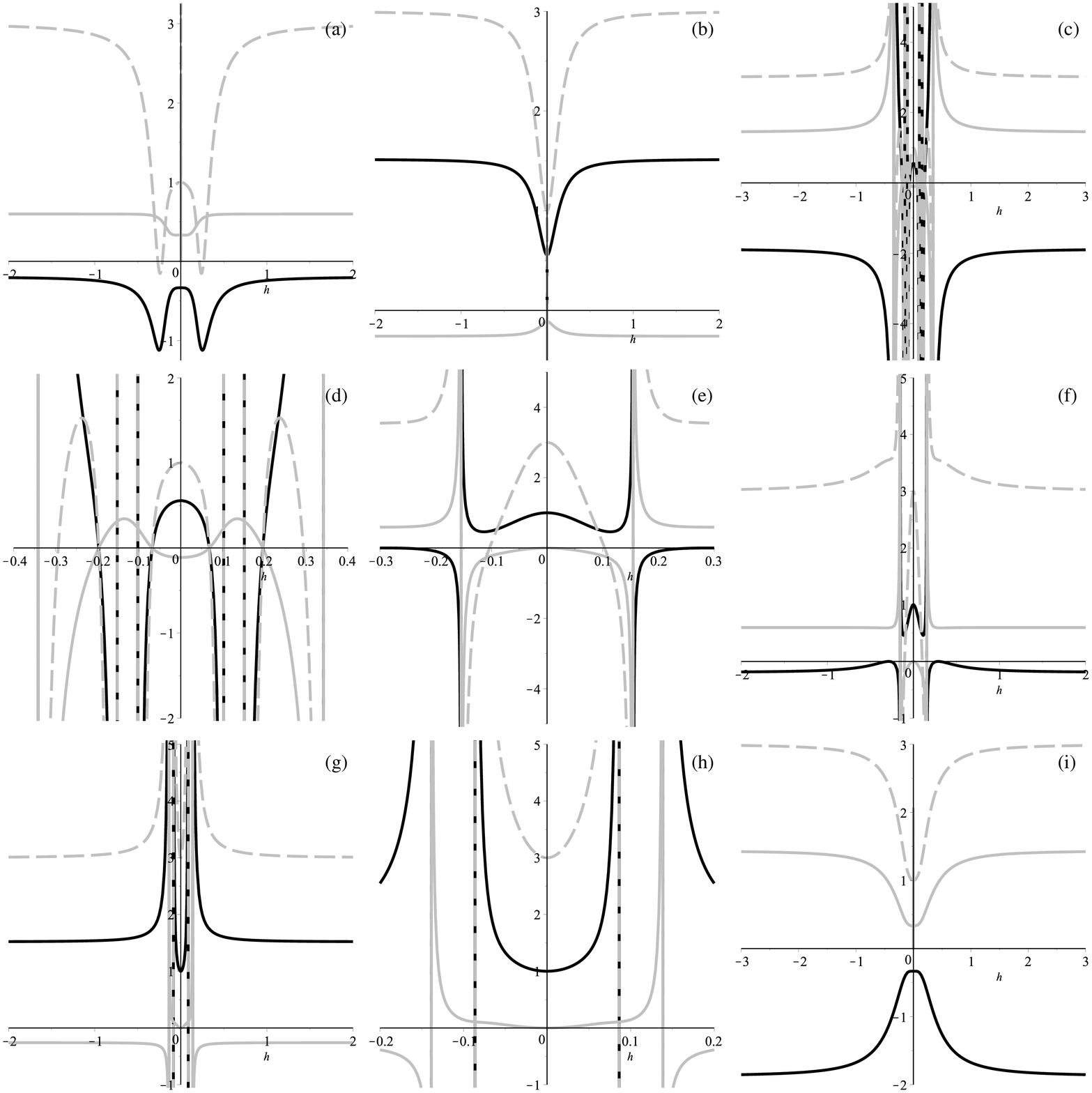}
\caption{Typical dynamics of Kasner exponents for $D\geqslant 4$ case. Panel (a) corresponds to $H_1$, $\alpha > 0$ case, panel (b) -- to $H_1$, $\alpha < 0$; panels (c) and (d) show $H_2$, $\alpha > 0$
at large scale (c) and fine structure in the vicinity of $h=0$ (d); (e) and (f) panels -- $H_2$ branch for $\alpha < 0$ with fine structure on (e) and large-scale structure on (f) panels; (g) and (h)
panels depict $H_3$ for $\alpha > 0$ -- large-scale on (g) and fine structure on (h); finally, (i) panel corresponds to $H_3$, $\alpha < 0$
(see text for details).}\label{D4_2}
\end{figure}

Similar to the previous sections, we also want to make an analysis in Kasner exponents -- ($p_H$, $p_h$) coordinates. So we define $p = - H^2/\dot H$ for both scale factors, use expressions for $\dot H$
and $\dot h$ and plot the resulting curves. They are presented in Fig.~\ref{D4_2} with the same definitions as in previous figures -- $p_H$ depicted by black line, $p_h$ -- by solid grey and $\sum p$
by dashed grey.
There (a) and (b) panels correspond to $H_1$ with $\alpha > 0$ in (a) and $\alpha < 0$ in (b); (c)--(f) panels depict
$H_2$: (c) and (d) are for $\alpha > 0$ with large-scale structure in (c) and fine structure in (d); (e) and (f) are for $\alpha < 0$ -- fine structure in (e) and large-scale structure in (f); finally,
panels (g)--(i) reflect $H_3$ branch -- (g) and (h) for $\alpha > 0$ -- large-scale and fine structures respectively and (i) is panel for $\alpha < 0$. Now let us analyze them and compare with Fig.~\ref{D4_1}.

Panel (a) of Fig.~\ref{D4_2} clearly demonstrate $K_1 \to K_3$ transition -- all according to Fig.~\ref{D4_1}(c). The same but in opposite ``direction'' ($h < 0$ in Fig.~\ref{D4_2}(a) has $p_h > 0$ while in
(b) panel we have $p_h < 0$ -- that is why the ``direction'' is reversed) we can see $K_3 \to K_1$ - again, all according to Fig.~\ref{D4_1}(g). Next two panels -- (c) and (d) -- represent $H_2$, $\alpha > 0$ case
-- we can see that the structure of regimes is as complicated as its counterpart in ($\dot H$, $\dot h$) coordinates (see Fig.~\ref{D4_1}(d)). We can depict $K_1$ at $h=0$ and after that nonstandard
singularity ($p_H,\,p_h \to 0$), which makes it $nS \to K_1$ transition -- minding $p_h < 0$ at that region. After that we detect another $nS$, making $nS \to nS$ transition, all according to Fig.~\ref{D4_1}(d).
Finally, after the second singularity we have vertical asymptote -- exponential solution -- and two surrounding regimes -- $E \to nS$ and $E \to K_3$. Panels (e) and (f) correspond to the same branch
$H_2$ but with $\alpha < 0$ and the corresponding dynamics in ($\dot H$, $\dot h$) is presented in Fig.~\ref{D4_1}(h). We can clearly see $K_3$ at $h=0$ and $h\to\infty$ as well as exponential solution
inbetween. According to the sign of $p_h$ in each region we can easily restore the behavior -- both regimes are $E \to K_3$, which agree with Fig.~\ref{D4_1}(h). Finally, $H_3$ branch is presented in the
bottom row of Fig.~\ref{D4_2} -- (g) and (h) panels correspond to $\alpha > 0$ while (i) -- to $\alpha < 0$. We can clearly see that regimes in (g) and (h) panels are ``reversed'' regimes in (e) and (f)
panels -- they are $K_3 \to E$. Finally in (i) panel we have $K_1 \to K_3$.

Now let us collect and list all regimes. Unlike previous sections we do not put conditions for them, just list them as they appear with growth of $h$. The results are presented in Table~\ref{D.4}. Also they
correspond only to regimes with $H>0$ so one of them ($H_2$, $\alpha < 0$) coming from $h>0$ part while others from $h<0$.

\begin{table}[h]
\begin{center}
\caption{Summary of $D\geqslant 4$ regimes with $H>0$.}
\label{D.4}
  \begin{tabular}{|c|c|c|}
    \hline
     Branch & $\alpha$   & Regimes  \\
    \hline
 \multirow{2}{*}{$H_1$} & $\alpha > 0$ & $K_1 \to K_3$ \\  \cline{2-3}
 & $\alpha < 0$ & $K_3 \to K_1$ \\  \cline{1-3}
  \multirow{5}{*}{$H_2$} &  \multirow{4}{*}{$\alpha > 0$} & $nS \to K_1$ \\   \cline{3-3}
  & & $nS \to nS$ \\ \cline{3-3}
  & & $E_{3+D} \to nS$ \\  \cline{3-3}
  & & $E_{3+D} \to K_3$ \\ \cline{2-3}
  & $\alpha < 0$ & $K_3 \to E_{iso}$ (both regimes) \\ \cline{1-3}
  \multirow{2}{*}{$H_3$} & $\alpha > 0$ & $K_3 \to E_{3+D}$ (both regimes) \\  \cline{2-3}
  & $\alpha < 0$ & $K_1 \to K_3$ \\
\hline
  \end{tabular}
\end{center}
\end{table}

To conclude, the choice of nonsingular late-time regimes in the general case is the same as in previous ones -- GR Kasner and exponential solutions -- either isotropic, or anisotropic. And with isotropic
solutions violate observations, we are left with GR Kasner and anisotropic exponential solutions. The former of them presented in $H_1$ branch at $\alpha < 0$ while the latter in $H_3$ branch
at~$\alpha > 0$.

\section{Discussions}

After collecting all the results, it is time to summarize and discuss them. Before turning to the results in each particular $D$, let us describe the similarities of all cases. First, all
asymptotic regimes are Kasner (GR or GB), exponential or singular. In the Introduction we mentioned that when looking for exact solutions, one usually consider either power-law or exponential {\it ansatz}
for scale factor and our research proved that it is absolutely right decision -- there are no nonsingular regimes in EGB cosmology apart from these two - at least with spatial splitting under consideration.
Both of these two regimes are already well-described (see Introduction for appropriate citations) and exponential solutions for our spatial splitting fall into two categories -- isotropic and anisotropic.
For lower dimensions ($D=1,\,2$) exponential solutions are described in~\cite{CPT1}, for higher dimensions ($D=3,\,4$) -- in~\cite{CST2}. From the results of~\cite{CPT1} we know that for 5D EGB model
(regardless of spatial splitting) there is only one stable vacuum exponential solution -- isotropic, and that is exactly what we obtained. Next, in 6D there are only two stable vacuum exponential
solutions --
isotropic and the solution with $(3+2)$ spatial splitting (i.e. spatial metric symmetry is given by $\{a_1,\,a_1,\,a_1,\,a_2,\,a_2\}$ -- product of three-dimensional and two-dimensional isotropic subspaces). Again,
this is exactly what we observe and the ratio of the Hubble parameters is in agreement with~\cite{CST2}. Next, 7D EGB model has~\cite{CST2} much more solutions but only two of them fit our $(3+D)$ spatial
splitting -- isotropic and $(3+3)$ one. The latter has two branches and one of them is stable for $H > 0$ while the other -- for $H < 0$. And again this is exactly what we see -- as we work with $H > 0$
only, we detect one anisotropic solution to be stable and the other -- unstable, plus isotropic solution. The same pattern -- stable isotropic plus anisotropic with two branches -- stable for $H>0$ and
stable for $H<0$ -- detected for $D=4$ (see~\cite{CST2}) and judging from the results of current paper, all $D \geqslant 4$ cases share the same pattern.

Let us start with $D=1$ case -- we have found that in this case there is only one viable regime -- $K_3 \to K_1$ transition which happens for $\alpha > 0$ regardless of the initial conditions. Regimes at
$\alpha < 0$ are either singular or exponential isotropic, which contradict observational data.
The next case -- $D=2$ -- has a bit more complicated structure. Indeed, unlike $D=1$ case where we have only one branch of solutions, $D=2$ case has two (see Eq.~\ref{D2_hvac}). So in this case we have two
viable regimes -- one of them is Kasner transition $K_3 \to K_1$ and it happens on $h_-$ branch regardless of $\alpha$ and initial conditions. The second one is an anisotropic exponential solution $E_{3+2}$
with expanding three-dimensional space ($H>0$) and contracting extra dimensions ($h<0$). It takes place only in $h_+$ branch and there is a lower bound on the 4D expansion rate (see Table~\ref{D.2}).
Actually, all cases started from this one have the same viable regimes but their distribution and the prerequisites are different. Indeed, $D=3$ case has both Kasner transition on $h_2$ branch with $\alpha < 0$ and
anisotropic exponential solution -- on the same branch but with $\alpha > 0$. Starting from $D \geqslant 4$ Kasner transition occurs on $H_1$ with $\alpha < 0$ while anisotropic exponential expansion --
on $H_3$ with $\alpha > 0$; isotropic exponential solution is on $H_2$ and with $\alpha < 0$ -- so that for $D \geqslant 4$ different exponential solutions resides on different branches plus stable Kasner is
located on third branch.

Just for comparison and without derivation (results obviously converge and the technics is the same as in Section~\ref{S3}) we can provide the results for $D=1$ and $D=2$ if we solve constraint equation
not for $h$, as we did in the corresponding sections, but for $H$, as we did for $D \geqslant 4$ case. This could be useful for understanding of the structure of solution and its variation with varying
number of extra dimensions.
In the $D=1$ case cubic equation reduced to quadratic so there are only two branches while in $D=2$ there are three. We put these results into the first two columns of Table~\ref{D.add}.

\begin{table}[h]
\begin{center}
\caption{Summary of nonsingular regimes.}
\label{D.add}
  \begin{tabular}{|c|c||c|c||c|c||c|c|}
    \hline
\multicolumn{2}{|c||}{$D=1$} & \multicolumn{2}{c||}{$D=2$} &  \multicolumn{2}{c||}{$D=3$} &  \multicolumn{2}{c|}{$D\geqslant 4$}\\
\hline
\multirow{3}{*}{$H_+$} & $K_3 \to K_1$ for $\alpha > 0$ & $H_1$ & $K_3 \to K_1$ for $\alpha < 0$ & \multirow{2}{*}{$H_1$} &  \multirow{2}{*}{$K_3 \to E_{iso}$ for $\alpha < 0$} & $H_1$  & $K_3 \to K_1$ for $\alpha < 0$\\ \cline{3-4} \cline{7-8}
 &  & \multirow{2}{*}{$H_2$} & $K_3 \to K_1$ for $\alpha > 0$ & &  & \multirow{2}{*}{$H_2$} & \multirow{2}{*}{$K_3 \to E_{iso}$ for $\alpha < 0$} \\ \cline{4-6}
 & $K_3 \to E_{iso}$ for $\alpha < 0$ & & $K_3 \to E_{iso}$ for $\alpha < 0$ & \multirow{3}{*}{$H_2$} &  $K_3 \to E_{3+3}$ for $\alpha > 0$ & & \\ \cline{1-4} \cline{7-8}
\multirow{2}{*}{$H_-$} & \multirow{2}{*}{$K_3 \to K_1$ for $\alpha > 0$} & \multirow{2}{*}{$H_3$} & $K_3 \to K_1$ for $\alpha > 0$  & & & \multirow{2}{*}{$H_3$} & \multirow{2}{*}{$K_3 \to E_{3+D}$ for $\alpha > 0$}\\ \cline{4-4}
 & & & $K_3 \to E_{3+2}$ for $\alpha > 0$  & & $K_3 \to K_1$ for $\alpha < 0$ & &\\
\hline
  \end{tabular}
\end{center}
\end{table}

One can see that in $D=1$ Kasner transition exists in both branches and in both with $\alpha > 0$ while isotropic solution in one of them and with $\alpha < 0$. In $D=2$ Kasner transition exist in all three
branches and with both signs; isotropic exponential solution exists in one of the branches with $\alpha < 0$ while anisotropic exponential solution -- in another branch with $\alpha > 0$. We can see
that in this approach different exponential solutions are not ``mixing'' inside the same branch, unlike approach we used in the main text -- for $D=2$ with ``usual'' approach both exponential solutions
exist on the same $h_+$ branch (see Table~\ref{D.2}). Also for  illustration purposes we summarized all nonsingular regimes in Table~\ref{D.add}. The $D=3$ case has the same description whenever we solve
constraint equation (\ref{D3_con}) with respect to $H$ or $h$ -- Eq. (\ref{D3_con}) is symmetric with respect to them, so we can just interchange $H \leftrightarrow h$ and get the same result.

So let us look closer on Table~\ref{D.add}. We can see that Kasner transitions $K_3 \to K_1$ in low dimensions occur at $\alpha > 0$ ($D=1$), then for both signs ($D=2$) for $\alpha$ and then for
$D \geqslant 3$ only for $\alpha < 0$. This is the effect of more complicated dynamics in higher number of dimensions.
Exponential solutions are always separated between branches -- two different exponential solutions do not exist on the same branch. Also, exponential solutions, their abundance and stability are in exact
agreement with the results of~\cite{CPT1, CPT3, my15}.

Of special interest are nonstandard singularities. This referred to a situation when some of the dynamical variables diverge while others are regular, at the same time curvature invariants diverge so it is a physical
singularity (say, ``suddenly'', at some regular value of scale factor, curvature invariants diverge). This kind of singularity is ``weak'' by Tipler's classification~\cite{Tipler}, and ``type II''
in classification by Kitaura and Wheeler~\cite{KW1, KW2}. Recent studies of the singularities of this kind in the cosmological context in Lovelock and Einstein-Gauss-Bonnet gravity
demonstrates~\cite{CGP2, mpla09, grg10, KPT, prd10} that their presence is not suppressed and they are abundant for a wide range of initial conditions and parameters. This is especially true for
Bianchi-I-type (i.e. where all scale factors are different -- $\diag(-1, a(t)^2, b(t)^2, c(t)^2, d(t)^2)$) (4+1)-dimensional EGB model~\cite{prd10} where it was demonstrated that in vacuum case recollapse
and nonstandard singularities are the only options for future behavior.

Before concluding our results, two important notes regarding the viability of the regimes must be done. First of them regards GR Kasner regimes.
We have found that $D=1$ GR Kasner regime has $p_H = 0.5$, $D=2$ case -- $p_H = \dac{1}{2\sqrt{6}-3} \approx 0.5266$, and for $D=3$ it is detected that $p_H = \dac{2}{3\sqrt{5}-3} \approx 0.5294$; further, for general $D \geqslant 4$ we derived

\begin{equation}
\begin{array}{l}
p_H = \dac{1}{3} - \dac{D + \sqrt{3D^2 + 6D}}{3(D+3)}~\mbox{with}~\lim\limits_{D\to\infty} p_H = \dac{1}{\sqrt{3}}\approx 0.577.
\end{array} \label{D4_pH}
\end{equation}

\noindent One can see that the resulting Kasner exponent $p_H$ in $K_1$ gradually grows from $0.5$ till approximately $0.577$, and that is Kasner exponent which is detected by an observer living in
$(3+1)$-dimensional space-time (``our Universe''). Transfering Kasner exponents to the expansion rate and remembering how scale factor depends on the equation of state in presence of the perfect fluid,  
we can write down
$p = 2/(3(1+\omega_{eff}))$, so $\omega_{eff} = 2/(3p) - 1$ -- effective equation of state which corresponds to the expansion rate. One can clearly see that for $D=1$ it is radiation ($\omega_{eff} = 1/3$)
and it becoming softer in the limit $D\to\infty$: $\omega_{eff} = 2/\sqrt{3}-1 \approx 0.1547$. We can see that all of them are different from $0$ what one would assume from dust-dominated Friedmann stage, 
so that
dust-dominated Friedmann behavior cannot be restored (unlike~\cite{CGPT} where it could).

The second note regards both Kasner and exponential solutions and it is linked to the Gauss-Bonnet coupling constant $\alpha$. As we demonstrated, Kasner solutions exist for $\alpha > 0$ at low $D$ and for 
$\alpha < 0$
at high $D$; isotropic exponential solutions exist only for $\alpha < 0$ while anisotropic -- only for $\alpha > 0$. So that if there were a bounds on $\alpha$, we could reject some of the regimes. And the
situation with bounds on $\alpha$ are the following -- from consideration of shear viscosity to entropy ratio as well as casuality violations and CFTs in dual gravity description there were obtained limits
on $\alpha$ for 5D that $\alpha/2 \leqslant 9/100$~\cite{alpha_01, alpha_02} and $\alpha/2 \geqslant -7/36$~\cite{alpha_03, alpha_04}; later they were updated for 7D~\cite{alpha_05, alpha_06, alpha_07}
$-5/16 \leqslant \alpha/2 \leqslant 3/16$ and eventually for any $D$~\cite{alpha_06, alpha_07} (with the upper limit found earlier in~\cite{alpha_08}):

\begin{equation}
\begin{array}{l}
- \dac{(3D + 11)(D+1)}{4(D+5)^2} \leqslant \dac{\alpha}{2} \leqslant \dac{D(D+1)(D^2 + 5D + 24)}{4(D^2 + 3D + 26)^2}.
\end{array} \label{alpha_limit}
\end{equation}

\noindent From these constraints one can clearly see that we cannot abandon either $\alpha > 0$ or $\alpha < 0$; study of GB superconductors~\cite{alpha_09} also do not allow to discard either possibility.
Considering black holes instabilities in 5D allow to lower upper limit to $\alpha < 1/24$~\cite{alpha_10} but it is still remains positive while in~\cite{alpha_11} $\alpha > 0$ brought some instabilities.
Overall, from all these studies we cannot disregard neither $\alpha > 0$ nor $\alpha < 0$. On the other hand, if we consider heterotic strings setup and so identify $\alpha$ with inverse string
tension (see~\cite{alpha_12}), then we should use $\alpha > 0$ only. In that case (if we restrain us with $\alpha > 0$ only) the only regimes which are viable are: Kasner transitions $K_3 \to K_1$ in 
$D=1,\,2$ and Kasner to anisotropic exponential transitions $K_3 \to E_{3+D}$ in $D \geqslant 2$.

\section{Conclusions}

It is time to draw conclusions to the results of this paper. As we mention in the Introduction, when looking for the exact solutions in EGB and Lovelock gravity, usually power-law and exponential 
{\it ansatz}
are considered. We have demonstrated that this approach is just -- at least in EGB case and at least when looking for the solutions which allow dynamical compactification. Indeed, we have demonstrated
analytically that there are no nonsingular regimes apart from power-law and exponential. And of power-law regimes, only Kasner regime is achieved (see~\cite{PT} for (un)viability of another power-law
regime). As of exponential regimes, their appearance, abundance and stability is in total agreement with theoretical predictions~\cite{CPT1, CPT3, my15}.

We have described distribution of all nonsingular regimes over branches and initial conditions. As we already mentioned, there are only two viable regimes -- $K_3 \to K_1$ Kasner regime and $K_3 \to E_{3+D}$
Kasner-to-exponential transition. The former of them occur for $\alpha > 0$ at low $D \leqslant 2$ and for $\alpha < 0$ at high $D \geqslant 2$ (at $D=2$ it exist for both sings of $\alpha$). 
On the contrary,
anisotropic exponential solution stable for $H > 0$ and $h<0$ (it corresponds to the expansion of our (3+1)-dimensional Universe -- that is what we observe and contraction of extra dimensions -- we do not 
sense them) only for $\alpha > 0$. 

To summarize the regimes, for
$\alpha > 0$ we have anisotropic exponential solutions at $D \geqslant 2$ (there is no such solution for $D=1$) as well as Kasner regime at $D=1,\,2$. On the contrary, for $\alpha < 0$ we have only Kasner 
regime
for $D \geqslant 2$ and so there is no viable regime for $D=1$. 
As we discussed earlier, current limits on $\alpha$ allow both signs so one needs to use additional reasoning to set
either $\alpha > 0$ (like appealing to inverse string tension in heterotic string theories) or $\alpha < 0$.

As we mentioned in the beginning, this is the first paper of the series, and we are about to finish similar paper but with boundary (cosmological, or $\Lambda$) term taken into account. Indeed, 
$\Lambda$-term case is broader and probably could offer more abundant dynamics, not to mention that one of the interpretations of current accelerated
expansion of the Universe is the $\Lambda$-term.
In future we are also going to consider matter in the form of the perfect fluid and probably the curvature of the manifolds.
The former of them could change late-time asymptote for Kasner regime -- we demonstrated that depending on $D$
late-time power-law regime $a(t) \propto t^p$ has $0.5 \leqslant p \leqslant 1/\sqrt{3} \approx 0.577$, which
contradict observations. So we hope that the addition of matter in form of the perfect fluid could change this asymptote
to favored by observations. As of the case with curved manifolds, we have considered that case numerically in~\cite{CGP1, CGP2, CGPT} but analytical consideration is always more reliable.

\begin{acknowledgments}
This work was supported by FAPEMA.
\end{acknowledgments}

\end{document}